\documentclass[a4paper,11pt]{article}
\pdfoutput=1 

\usepackage{jcappub} 

\usepackage[T1]{fontenc} 
\usepackage[sort&compress]{natbib}
\usepackage[utf8]{inputenc}
\usepackage{multirow}
\usepackage[english]{babel}
\usepackage{hyperref}
\usepackage{amsmath}
\usepackage{amssymb}
\usepackage{epsfig}
\usepackage{ccaption}
\usepackage{lscape}
\usepackage{footnotebackref}
\usepackage{longtable}
\usepackage{float}
\usepackage{graphics,psfrag,rotating}
\usepackage{graphicx}
\usepackage{dcolumn}
\usepackage{bm}
\usepackage{array,multirow}
\usepackage{xspace}
\usepackage{xcolor}
\usepackage{placeins}
\usepackage{afterpage}
\usepackage{etoolbox}
\usepackage{xcolor,colortbl}
\usepackage{array,longtable}
\definecolor{Gray}{gray}{0.85}

\def\arraystretch{1.25}

\begin{document}

\title{A robust estimate of the Milky Way mass from rotation curve data}

\author[a,b]{E.V.~Karukes,}
\author[b,c]{M.~Benito,}
\author[b,d,e]{F.~Iocco,}
\author[d,f,g]{R.~Trotta,}
\author[d]{A.~Geringer-Sameth}

\affiliation[a]{Astrocent, Nicolaus Copernicus Astronomical Center Polish Academy of Sciences, ul. Bartycka 18, 00-716 Warsaw, Poland}
\affiliation[b]{ICTP-SAIFR \& IFT-UNESP, R. Dr. Bento Teobaldo Ferraz 271, S\~ao Paulo, Brazil}
\affiliation[c]{National Institute of Chemical Physics and Biophysics, R\"avala 10, Tallinn 10143, Estonia}
\affiliation[d]{Physics Department, Astrophysics Group, Imperial Centre for Inference and Cosmology, Blackett Laboratory, Imperial College London, Prince Consort Rd, London SW7 2AZ}
\affiliation[e]{Universit\`a di Napoli ``Federico II'' \& INFN Sezione di Napoli, Complesso Universitario di Monte S.~Angelo, via Cintia, 80126 Napoli, Italy}
\affiliation[f]{Data Science Institute, William Penney Laboratory, Imperial College London, London SW7 2AZ}
\affiliation[g]{SISSA, Data Science Excellence Department, Via Bonomea 265, 34136 Trieste, Italy}

\emailAdd{ekarukes@camk.edu.pl}
\emailAdd{mariabenitocst@gmail.com}
\emailAdd{fabio.iocco.astro@gmail.com}
\emailAdd{r.trotta@imperial.ac.uk}
\emailAdd{a.geringer-sameth@imperial.ac.uk}

\newcommand{\FI}[1]{{\color{blue}[{\bf FI:} #1]}}
\newcommand{\EK}[1]{{\color{red}[{\bf EK:} #1]}}
\newcommand{\MB}[1]{{\color{orange}[{\bf MB:} #1]}}
\newcommand{\RT}[1]{{\color{magenta}[{\bf RT:} #1]}}
\newcommand{\UNI}[1]{{\color{purple} #1}}
\newcommand{\rt}[1]{\RT{#1}}
\newcommand{\AGS}[1]{{\color{cyan}[{\bf AGS:} #1]}}
\newcommand{\sigmaint}{\sigma_\text{int}}

 \abstract{We present a new estimate of the mass of the Milky Way, inferred via a Bayesian approach by making use of tracers of the circular velocity in the disk plane and stars in the stellar halo, as from the publicly available {\tt galkin} compilation. We use the rotation curve method to determine the dark matter distribution and total mass under different assumptions for the dark matter profile, while the total stellar mass is constrained by surface stellar density and microlensing measurements. We also include uncertainties on the baryonic morphology via Bayesian model averaging, thus converting a potential source of systematic error into a more manageable statistical uncertainty. 
 We evaluate the robustness of our result against various possible systematics, including rotation curve data selection, uncertainty on the Sun's velocity $V_0$, dependence on the dark matter profile assumptions, and choice of priors. We find the Milky Way's dark matter virial mass to be 
 $\log_{10}M_{200}^{\rm DM}/ {\rm M_\odot} = 11.92^{+0.06}_{-0.05}{\rm(stat)}\pm{0.28}\pm0.27{\rm(syst)}$ ($M_{200}^{\rm DM}=8.3^{+1.2}_{-0.9}{\rm(stat)}\times10^{11}\,{\rm M_\odot}$).
We also apply our framework to Gaia DR2 rotation curve data and find good statistical agreement with the above results.}

\maketitle
\flushbottom
\section{Introduction}
\label{sec:intro}

In the standard cosmological paradigm, only $\sim15 \% $ of the total matter density in the Universe is in the form of ordinary matter, while dark matter makes up the other $85 \%$~\cite{Ade:2015xua}. The existence of dark matter has long been inferred from its gravitational interactions with ordinary luminous matter on scales ranging from galaxies to the Universe as a whole (for reviews see e.g.~\cite{Bertone:2010zza,Buckley:2017ijx}). Observations of the Universe on large scales are accurately described by the concordance cosmological model, known as $\Lambda$CDM, which includes cold dark matter along with a cosmological constant. While $\Lambda$CDM successfully describes the observed large scale structure and dynamics, some observational discrepancies seem inconsistent with its predictions on small scales (see e.g.~\cite{deblok02,gentile05,boylan-kolchin12,Karukes:2016eiz}. 
The central question is whether these discrepancies arise from our inability to accurately model complex but known physical processes or whether they represent a fundamental inadequacy in the standard paradigm. Indeed, in recent years high-resolution hydrodynamical simulations that self-consistently take into account baryonic feedback indicate that the small-scale discrepancies can largely be mitigated within the $\Lambda$CDM model~\cite{Sawala:2016tlo}. 
In this scenario, various discrepancies, such as the missing satellites~\cite{moore99,Klypin+1999} and the too-big-to-fail problems~\cite{BoylanKolchin+2011}, strongly depend on the assumed Milky Way mass, which, if estimated incorrectly, may lead to biased conclusions (e.g~\cite{Purcell:2012kd,Wang_2012,Vera_Ciro_2012,Cautun+2014}). In addition, tests of alternative warm dark matter models~\cite{Kennedy:2013uta,Lovell:2013ola} also require knowledge of the total halo mass.  Thus, the Milky Way's total mass and the mass of its dark matter halo are quantities of particular interest, because they enable certain tests of the current cosmological model~\cite{Smith:2006ym,2013ApJ...768..140B,2014A&A...562A..91P}. Surprisingly, despite being a consequential parameter, the total mass of the Milky Way is poorly constrained. Therefore, it is crucial to be able to put stringent constraints on the Milky Way mass, which compliment other mass estimates from the existing literature and also account for different systematic errors.

There are various techniques used to constrain the mass of the Galaxy. Each have their advantages and shortcomings and are affected by different sources of systematic error (see~\cite{BlandHawthorn&Ortwin2016} for a review). Rather than relying on a particular technique and measurement, it is important to estimate the total Milky Way mass using different methods.
Despite much effort, the mass of the Galaxy currently carries a factor of four uncertainty. Even considering only the most recent studies using Gaia data, the inferred Milky Way halo mass ranges from $M_{200}^{\rm DM}=(6-22)\times 10^{11} \,\mathrm{M}_{\odot}$, \footnote{There is no unique convention to define a galaxy's halo mass (see e.g.~\cite{Bryan&Norman1998, Klypin+2002, Klypin+2011}). In this work, we define the halo mass $M_{200}^{\rm DM}$ as the mass of dark matter enclosed within a sphere which has an average density $200$ times the critical density of the Universe.
}~\cite{Eadie&Juric2018, Monari+2018, 2019A&A...621A..56P, Deason+2019, Callingham+2019}.

This work builds on the rotation curve analysis presented in~\cite{Karukes:2019jxv}, hereafter called Paper~I. The aim of this paper is to provide a determination of the total mass of the Milky Way and of its dark matter component. We demonstrate that our results provide precise and accurate constraints, while being robust to various systematic uncertainties. 
Our results are compatible with the most recent estimates using other techniques and our method can easily incorporate new data sets over the entire range of galactocentric distances we consider in our study.

The paper is structured as follows:
in Section~\ref{sec:Methods} we describe the astrophysical data sets used for the mass determination and the statistical procedures we adopt. In Section~\ref{sec:results} we present our results for our fiducial astrophysical setup.

In Section~\ref{sec:tests} we carry out tests of robustness using both mock data as well as by considering various systematic uncertainties, and varying our astrophysical setup. In Section \ref{sec:compmass} we compare our results with other estimates in the literature, and also apply our own procedure to the Gaia DR--2 data.
We conclude in Section~\ref{sec:conclusions}.

\section{Methodology and Data}
\label{sec:Methods}
In this work, we further develop the methodology presented in Paper~I. We analyse the observed galactic rotation curve in a Bayesian framework in order to constrain a model describing both the large-scale distribution of baryons as well as the dark matter halo. We then marginalize the resulting posterior probability distribution over the baryonic and dark components to obtain a determination of the Milky Way's total mass.

There are two main differences with respect to the analysis presented in Paper~I. First, we consider an additional prior distribution for the dark matter halo parameters (Section~\ref{sec:DM}) to verify the robustness of our results with respect to choice of priors. Second, we employ Bayesian model averaging to include a range of various possible baryonic morphologies (Section~\ref{subsubsec:model_averaging}). As a result of this averaging procedure, our estimate of the Milky Way mass fully includes systematic uncertainties arising from our ignorance of the exact shape of the baryonic distributions. 

The structure of this section is as follows: in Sections~\ref{sec:data sets} and~\ref{sec:baryons} we briefly describe the rotation curve observations and the various baryonic mass distributions. 
Section~\ref{sec:DM} describes the model of the Milky Way's dark matter halo. Finally, the statistical framework is described in Section~\ref{sec:statistic}.
We refer the reader to Paper~I and references therein for a detailed description of the astrophysical setup (observations of the rotation curve, and of the luminous component of the Galaxy) and statistical framework adopted in this work.

\subsection{The observed rotation curve}
\label{sec:data sets}
We adopt two different compilations of Milky Way rotation curve observations, the {\tt galkin}~\cite{Pato:2017yai} compilation and that of Huang~{\it et~al.}~\cite{2016MNRAS.463.2623H}. 
The {\tt galkin} compilation consists of 25 data sets that comprise a number of different kinematic tracers (gas, stars, and masers) of the total gravitational potential within the visible Galaxy. Measurements extend to galactocentric distances of $\sim 25$~kpc. The Huang~{\it et~al.}~\cite{2016MNRAS.463.2623H} compilation consists of two data sets (hereafter referred as Huang$_1$ and Huang$_2$), probing the total gravitational potential up to $\sim 15$~kpc and $\sim 100$~kpc, respectively. Notice that {\tt galkin} and the Huang~{\it et~al.} data sets overlap between 8 and 20~kpc.

We start by fixing the Sun's distance to the galactic Centre to $R_0=8.34$~kpc and its circular velocity to $V_0=239.89$~km/s~\cite{2016MNRAS.463.2623H}. For the peculiar motion of the Sun, we adopt $(U_{\odot},V_{\odot},W_{\odot})=(7.01, 12.20, 4.95)$~km/s \cite{2016MNRAS.463.2623H}. 
This choice corresponds to the one made by Huang~{\it et~al.}~\cite{2016MNRAS.463.2623H} and is necessary in order to combine Huang$_1$ and Huang$_2$ with {\tt galkin}. In Section~\ref{subsec:V0} we explore the robustness of our results when modifying various assumptions, including the galactic parameters ($R_0$,$V_0$).

In Paper~1 we presented a method based on Bayesian model comparison to identify a mutually compatible subset of the {\tt galkin} data we call \texttt{galkin$_{12}$}. In summary, the method uses the Bayesian evidence from different combinations of data sets as a discriminant to determine which sets are mutually compatible. Data sets that are in systematic tension with the rest of the data are discarded, in order to avoid biasing subsequent inference. With this procedure, out of the 25 data sets of the {\tt galkin} compilation we select a subset of 12 mutually compatible data sets, which are then binned in exactly the same manner as in Paper~I. We use this resulting \texttt{galkin$_{12}$} data set for the present analysis.

\subsection{The visible (baryonic) component}
\label{sec:baryons}

The exact distribution of baryons within the Galaxy is currently still debated, e.g.~\cite{BlandHawthorn&Ortwin2016}. In order to cope with this uncertainty, we adopt a large array of three-dimensional density profiles --- motivated by observations --- to describe the mass distributions of three baryonic components of the Galaxy: stellar bulge, stellar disk, and gas. By considering every permutation of baryonic profiles for the components we obtain a set of possible morphologies which bracket the systematic uncertainty on the distribution of the baryonic mass in our Galaxy, an approach first adopted in~\cite{Iocco:2015xga} and then followed by \cite{, 2015JCAP...12..001P,Karukes:2019jxv}.

Following the approach of~\cite{Iocco:2015xga}, we combine disks and bulges individually in order to remain agnostic as to their relative viability. That is, we express no preference on which bulge and disk models are preferred, but present results properly averaged (see Sec.~\ref{subsubsec:model_averaging}) over all possible combinations.
For the gas component, we keep the shape of the morphology and total mass fixed as its contribution to the gravitation potential is subleading and including its uncertainty would not affect our results~\cite{Iocco:2015xga, 2015JCAP...12..001P}.
Each baryonic morphology is named by using an abbreviation specifying the bulge followed by one specifying the disk. For example, the model G2BR is a combination of bulge profile G2~\cite{Stanek:1995ws} and disk profile BR~\cite{2013ApJ...779..115B}. We also present a summary of the morphologies we consider in Appendix~\ref{App:baryonic_morphologies} and table~\ref{tab:virial_radius_mass}.

Besides morphology, the total mass within each baryonic component is another source of uncertainty. In order to account for these uncertainties, we normalise the stellar disk profile by a parameter $\Sigma_*$ that sets the stellar surface density at the Sun's position~\cite{2013ApJ...779..115B} and we normalize the bulge mass using the microlensing optical depth towards the galactic center $\langle\tau\rangle$~\cite{Popowski:2004uv}. Both $\Sigma_*$ and $\langle\tau\rangle$ are then fitted to the observations alongside all other free parameters in the model, with a prior determined by the observational constraints on these quantities (see Section~\ref{sec:priors}). This procedure is thoroughly described in Paper~I, and we refer the reader to it and references therein for further details. We note that it is straightforward to include additional observations which constrain combinations of baryonic morphologies by adding terms to the likelihood (Eq.~\ref{eqn:L}) analogous to those describing microlensing optical depth and local stellar surface density.

\subsection{The dark matter halo}
\label{sec:DM}
The density of dark matter as a function of galactocentric radius $r$ can be modelled by a spherical generalized  Navarro, Frenk, and White (gNFW) profile~\cite{Zhao:1995cp,2001ApJ...555..504W}:

\begin{equation}\label{eqn:gNFW}
 \rho_\mathrm{gNFW}(r;r_s,\rho_s,\gamma)=\frac{\rho_s}{\left(\frac{r}{r_s}\right)^{\gamma}\left(1+\frac{r}{r_s}\right)^{3-\gamma}},
\end{equation}
where $r_s$ is the characteristic radius of the halo, $\rho_s$ is the characteristic dark matter density, and $\gamma$ is the logarithmic slope of the inner density profile. The value $\gamma=1$ corresponds to the standard NFW profile.

In order to estimate the Milky Way mass, we rewrite Eq.~\eqref{eqn:gNFW} in terms of the virial mass $M_\mathrm{vir} \equiv M(< R_\mathrm{vir})$ and the concentration $c \equiv R_\mathrm{vir}/r_{-2}$. The virial radius $R_\mathrm{vir}$ is the radius of the sphere in which the average dark matter density equals $\Delta$ times the critical density of the Universe\footnote{In this work we adopt a critical density of $\rho_{cr}=9.1\times10^{-30}\,\rm{g/cm^3}$ \cite{2013ApJS..208...19H}.} $\rho_{cr}$, while $r_{-2}$ is the radius at which the logarithmic slope of the density profile ($d\ln \rho / d\ln r$) is $-2$, which for a gNFW halo occurs at $r_{-2}=(2-\gamma)r_s$. There is no agreed unique choice for $\Delta$ (see e.g. \cite{Bryan&Norman1998, Klypin+2002, Klypin+2011}) and here we adopt $\Delta=200$. We relabel, accordingly, the virial radius and the virial mass as $R_{200}$ and $M_{200}^{\rm DM}$. With these definitions in hand, the relation between virial radius and virial mass is
\begin{equation}
    M_{200}^{\rm DM} = \frac{4\pi}{3}\,200\,\rho_{cr}\,R_{200}^3,
\end{equation}
while in terms of the gNFW profile we have
\begin{equation}\label{eqn:mass gNFW}
\begin{split}
     M_{200}^{\rm DM} &= \int\limits_0^{R_{200}} \rho_\mathrm{gNFW}(r) 4\pi r^2 dr \\
      & = -4\pi R_{200}^3 \left(\frac{R_{200}}{r_s}\right)^{-\gamma}\rho_s \frac{_2F_1[3-\gamma,3-\gamma;4-\gamma;-\frac{R_{200}}{r_s}]}{3-\gamma},
     \end{split}
\end{equation}

\noindent
where $_2F_1(a,b;c;z)$ is the ordinary hypergeometric function. Equating these two expressions for $M_{200}^{\rm DM}$ and using the definition of the concentration parameter yields an expression for $\rho_s$ in terms of $c$ and $\gamma$:

 \begin{equation}\label{eqn:scale density}
     \rho_s=\frac{200}{3}\rho_{cr}\frac{3-\gamma}{\left(c\,(2-\gamma)\right)^{-\gamma}\,_2F_1[3-\gamma,3-\gamma,4-\gamma,-c\,(2-\gamma)]}
 \end{equation}

By combining Eq.~\eqref{eqn:scale density} with the definitions of the scale radius and the virial radius, Eq.~\eqref{eqn:gNFW} can be expressed in terms of $c$, $M_{200}^{\rm DM}$, and $\gamma$. Then, by integrating the gNFW dark matter density we can obtain the dark matter mass enclosed within a given radius $M_\mathrm{DM}(<r)$. 

In Section~\ref{sec:tests} we consider two other dark matter density profiles: the Einasto~\cite{1965TrAlm...5...87E} and the Burkert~\cite{1995ApJ...447L..25B} profiles. The Einasto profile can be expressed as

\begin{equation}
    \rho_\mathrm{Ein}(r)=\rho_{-2}\,{\rm exp}\left\{-\frac{2}{\alpha}\left(\left(\frac{r}{r_{-2}}\right)^\alpha-1\right)\right\},
\label{eqn:Einasto}
\end{equation}
\noindent
where $\rho_{-2}$ and $r_{-2}$ are the density and radius at which $\rho(r)\propto r^{-2}$, and $\alpha$ is the Einasto index which determines the shape of the profile, yielding a core towards the central regions of a galaxy when $\alpha \gtrsim 1$. The Burkert profile can be written as
\begin{equation}
    \rho_\mathrm{Bur}(r) = \frac{\rho_0\,r_c^3}{(r+r_c)\,(r^2+r_c^2)},
\label{eqn:Burkert}    
\end{equation}
\noindent
where $\rho_0$ and $r_c$ are the core density and the core radius, respectively.

\subsection{Statistical framework}
\label{sec:statistic}

The observed rotation curve described in Section~\ref{sec:data sets} is governed by the total (baryonic + dark matter) distribution of mass in the Milky Way.
We use the rotation curve data, in combination with information on the distribution of gas, stars and dark matter as described in Section~\ref{sec:baryons}, to perform the global mass modeling and constrain the underlying dark matter distribution. To do so we fit a global model of the Galaxy that consists of four components:  stellar disk, gaseous disk, stellar bulge and dark matter halo. Each mass component contributes to the total circular rotation curve $\omega_\mathrm{tot}$ according to

\begin{equation}
\omega_\mathrm{tot}^2(\Theta, \Sigma_*, \langle\tau\rangle)=\omega_\mathrm{disk}^2(\Sigma_*)+\omega_\mathrm{gas}^2+\omega_\mathrm{bulge}^2(\Sigma_*, \langle\tau\rangle)+\omega^2_\mathrm{DM}(\Theta),
\label{eqn:vel_model}
\end{equation}
\noindent
where the first three baryonic components are described in Section~\ref{sec:baryons} and the dark matter component (see Section~\ref{sec:DM}) depends on parameters $\Theta = (c, M_{200}^{\rm DM}, \gamma)$. Note that each term in Eq.~\eqref{eqn:vel_model} is implicitly a function of galactocentric radius $r$ and that angular velocities $\omega_i$ are used instead of linear circular velocities ($V_i =  r\omega_i$).

\subsubsection{Priors and likelihood \label{sec:priors}}

 For a given baryonic morphology, our gNFW model has five free parameters: the concentration parameter of the dark matter halo $c$, the dark matter halo mass $M_{200}^{\rm DM}$, the logarithmic slope of the inner dark matter density profile $\gamma$, the microlensing optical depth $\left<\tau\right>$, and the stellar surface density at the Sun's position $\Sigma_*$. We work in a Bayesian framework which requires setting prior distributions on the model parameters. We adopt uniform priors over the following variables and ranges:

\begin{equation}\label{eqn:priors}
\begin{aligned}
    c &\in [0, 100],\\
    \log_{10}\frac{M_{200}^{\rm DM}}{\mathrm{M}_\odot} &\in [10,13],\\
    \gamma &\in [0.1,2],\\
    \frac{\Sigma_* }{ 10^7 \,\mathrm{M}_{\odot}\, \mathrm{kpc}^{-2}} &\in [1.9, 5.7],\\
   \frac{ \left<\tau\right>}{10^{-6}} &\in [0.1, 4.5].
\end{aligned}
\end{equation}

We use fairly wide priors, which encompass the support of the likelihood. The last two parameters ($\Sigma_*$ and $\left<\tau\right>$) are nuisance parameters which are each independently constrained by Gaussian likelihoods. For the means and standard deviations of these likelihood components we adopt the values of the stellar surface density at the Sun's position $R_0$ provided by~\cite{2013ApJ...779..115B}, $\Sigma_*^{\rm obs}=(3.8\,\pm\, 0.4)\times10^7\,\rm M_{\odot}/\rm kpc^2$, as well as the measurement of the microlensing optical depth provided by the MACHO collaboration in Popowski~et~al.~\cite{Popowski:2004uv}\footnote{We have explicitly checked that by using the most-recent MOA-II microlensing measurements (i.e. table~3 of~\cite{Sumi&Penny2016}) estimates of the Milky Way mass remain unchanged.}, $\langle\tau\rangle^{\rm obs}=2.17^{+0.47}_{-0.38}\times10^{-6}$. For simplicity, we symmetrize the error in the microlensing optical depth by adopting a standard deviation of $\sigma_{\langle\tau\rangle}=0.47$ which is conservative, as it uses the larger of the upper and lower error bar. 

The likelihood function is given in Eq.~(3.3) of Paper~I with the only difference being that in this analysis the dark matter distribution is parameterized by  $\Theta=(c, M_{200}^{\rm DM}, \gamma)$. We show in Section~\ref{subsec:twosearches} that changing the prior by adopting instead the set $\Theta=(\gamma, r_s, \rho_0)$ does not change our results appreciably. 

For a given choice of baryonic morphology, denoted by $\mathcal M$, the likelihood function takes the form:

\begin{equation}
\begin{split}
P({\rm d}|\Phi, \mathcal{M}) & = 
\prod_{i=1}^m\left\{\frac{1}{\sqrt{2\pi}\sigma_{\bar{\omega},i}}\exp\left[-\frac{1}{2}\frac{\left(\omega_c(r_i, \Phi)-\bar{\omega}_i\right)^2}{\sigma_{\bar{\omega},i}^2}\right]\right\} \\
&\times \frac{1}{\sqrt{2\pi}\sigma_{\langle\tau\rangle}}\exp{\left[-\frac{1}{2}\frac{\left(\langle\tau\rangle - \langle\tau\rangle^{\rm obs}\right)}{\sigma_{\langle\tau\rangle}^2}\right]} \\
& \times \frac{1}{\sqrt{2\pi}\sigma_{\Sigma_*}}\exp{\left[-\frac{1}{2}\frac{\left(\Sigma_* - \Sigma_*^{\rm obs}\right)}{\sigma_{\Sigma_*}^2}\right]},
\label{eqn:L}
\end{split}
\end{equation}

\noindent
where we have defined the parameter vector $\Phi=(c,M_{200}^{\rm DM},\gamma, \Sigma_*,\langle\tau\rangle)$, $\bar{\omega}_i$ is the measured angular velocity, $\sigma_{\bar{\omega},i}$ is the corresponding uncertainty, and $i$ runs over the radial rotation curve bins.
The posterior is obtained via Bayes theorem as 

\begin{equation}\label{eqn:Bayes}
P(\Phi|{\rm d,\mathcal{M}})=\frac{P({\rm d}|\Phi,\mathcal{M})P(\Phi|\mathcal{M})}{P(\rm{d}|\mathcal{M})},
\end{equation}
where $\mathcal{M}$ represents the assumed baryonic morphology (see Eq.~\eqref{eqn:vel_model}) and the likelihood
$P(\rm{d}|\Phi,\mathcal{M})$ is given by Eq.~\eqref{eqn:L}. The prior $P(\Phi|\mathcal{M})$ is separable in the model's parameters and is specified in Eq.~\eqref{eqn:priors}. The normalizing constant $P(\rm{d}|\mathcal{M})$ is called ``Bayesian evidence'' or ``model likelihood''.

\subsubsection{Bayesian model averaging}
\label{subsubsec:model_averaging}

Given the uncertainty in the choice of the baryonic morphology, we wish to incorporate this systematic uncertainty into our final Milky Way mass estimate. Bayesian model averaging (see e.g. \cite{Trotta_2008}) allows us to marginalize over the choice of baryonic morphology by treating an index specifying baryonic morphology type as an additional nuissance parameter. The procedure automatically downweights baryonic morphologies that are disfavoured by the rotation curve data, thus encapsulating an Occam's razor principle. 
This method has been successfully applied in various cosmological and astrophysical settings, see e.g. \cite{Marshall:2003ez,2011MNRAS.413L..91V,Palmese:2019lkh}. 

We denote each choice of baryonic morphology by $\mathcal{M}_i$. The model-averaged posterior for the parameters $\Phi$ is given by:
\begin{equation}\label{eqn:BMA}
\begin{aligned}
P(\Phi|{\rm d}) &= \sum_i P(\Phi, \mathcal{M}_i|{\rm d})
= \sum_i P(\Phi|{\rm d},\mathcal{M}_i) P(\mathcal{M}_i|{\rm d})\\
& =P(\mathcal{M}_0|{\rm d})\sum_i B_{i0} \frac{P(\mathcal{M}_i)}{P(\mathcal{M}_0)}P(\Phi|{\rm d},\mathcal{M}_i),
\end{aligned}
\end{equation}
where $i$ runs over all possible baryonic morphologies and $\mathcal{M}_0$ denotes an arbitrary reference morphology. Following~\cite{2017JCAP...02..007B}, we choose $\mathcal{M}_0=\rm E2HG$ (see table~\ref{tab:virial_radius_mass}), as it is the morphology that gives median rotation velocities with respect to all others. The Bayes factor $B_{i0}$ is the ratio of the Bayesian evidences between model $\mathcal{M}_0$ and model $\mathcal{M}_i$, obtained in each case by integrating the product of the likelihood and the parameters' prior over the entire parameter space:
\begin{equation}\label{eqn:Bayes factor}
B_{i0} \equiv \frac{P(\rm{d}|\mathcal{M}_i)}{P(\rm{d}|\mathcal{M}_0)} = \frac{\int d\Phi P(d|\Phi, \mathcal{M}_i)P(\Phi|\mathcal{M}_i)}{\int d\Phi P(d|\Phi, \mathcal{M}_0)P(\Phi|\mathcal{M}_0)}.
\end{equation}

If we assign equal prior probability to each of the $N=30$ baryonic morphologies we consider, i.e., $P(\mathcal{M}_i) = 1/N$ ($i=0,\dots, N-1)$, the prior ratio cancels in Eq.~\eqref{eqn:BMA}, and the expression for the model-averaged posterior becomes simply: 
\begin{equation}\label{eqn:BMA2}
P(\Phi|{\rm d}) \propto \sum_{i=0}^{N-1} B_{i0} P(\Phi|{\rm d},\mathcal{M}_i).
\end{equation}
In other words, we obtain the model-averaged posterior (up to an irrelevant constant) by taking the posterior samples from each baryonic morphology $i$ and weighing them according to the Bayes factor between model $i$ and the reference morphology. The model averaged posterior distribution then gives constraints on parameters $\Phi$ incorporating the additional uncertainty coming from the unknown shape of the baryonic components.

Finally, we notice that the priors in the evidence integral in Eq.~\eqref{eqn:Bayes factor} are identical for all the baryonic morphologies, i.e. $P(\Phi | \mathcal{M}_i) = P(\Phi | \mathcal{M}_0)$ for all $i$. Since the parameters' priors control the strength of the Occam's razor penalty for each model (see ~\cite{Trotta_2008} for details), we can be reassured that the penalty is the same for all baryonic morphologies. This introduces additional robustness in our model-averaged results: since the Bayes factor scales approximately linearly with the width of each prior in Eq.~\eqref{eqn:priors}, a change in the range for the uniform priors will translate into an approximate linear rescaling of each baryonic morphology's evidence, which cancels in the Bayes factor of Eq.~\eqref{eqn:Bayes factor}. Therefore we can conclude that the exact choice of prior range for the model parameters is unimportant for our model-averaged results (as long as the prior width is larger than the support of the likelihood, which is the case here). 

\subsubsection{Posterior sampling and evidence estimation}

We draw samples from the posterior distribution (conditional on a given baryonic morphology) by using the open source nested sampling code \texttt{PyMultiNest}~\cite{Buchner:2014nha}. \texttt{PyMultiNest} is a Python interface for \texttt{MultiNest}~\cite{Feroz:2008wr,Feroz:2008xx,Feroz:2013hea}, a generic Bayesian inference tool implementing the nested sampling algorithm~\cite{2004AIPC..735..395S}. The Bayesian model averaging analysis requires the calculation of the Bayesian evidence, which is 
the primary reason we use \texttt{MultiNest} instead of conventional Markov Chain Monte Carlo (MCMC). \texttt{PyMultiNest} delivers at the same time both posterior samples and an estimate of the Bayesian evidence, which we then use to compute the Bayes factor entering Eq.~\eqref{eqn:BMA2}.
We also perform an accuracy test against mock data (see Section~\ref{sec:accuracy}) and explore the effect of different choices of prior (Section~\ref{subsec:twosearches}) using the open source affine-invariant Markov Chain Monte Carlo (MCMC) ensemble sampler \texttt{emcee}~\cite{2013PASP..125..306F}. It is also used in some of the runs of Section~\ref{sec:tests} where the calculation of the Bayesian evidence is not required. As a further test of the numerical stability of our results, we have checked that we obtain identical results for the posterior distributions for a given morphology when using \texttt{PyMultiNest} and \texttt{emcee}, up to sampling noise.

\begin{table}
\centering
\addtolength{\tabcolsep}{1.5pt}
\scalebox{1}{%
\begin{tabular}{ | l | c | c | c | c | c | c |}
\hline 
Baryonic & $R_{200}$ & $c$ &$M_{200}^{\rm DM}$  &  $M_{\rm bar}$  &$M_{\rm tot}$ & $B_{i0}$\\
morphology  & [kpc]  &   & [$10^{11}\;\mathrm{M}_{\odot}$] & $[10^{10} \; {\mathrm{M}_{\odot}}]$ &  [$10^{11}\; \mathrm{M}_{\odot}$] & --- \\
\hline
\rowcolor{Gray}
G2~\cite{Stanek:1995ws}BR~\cite{2013ApJ...779..115B}  & $201^{+8}_{-5}$ &  $16^{+2}_{-2}$ & $9.3^{+1.2}_{-0.8}$ & $6.7^{+0.4}_{-0.3}$ & $10.0 ^{+1.0}_{-0.7}$ & 0.24\\
E2~\cite{Stanek:1995ws}BR~\cite{2013ApJ...779..115B}  & $200^{+8}_{-6}$ &  $16^{+1}_{-2}$ & $9.2^{+1.1}_{-0.7}$ & $6.8^{+0.3}_{-0.4}$& $9.9 ^{+1.0}_{-0.6}$ & 0.21\\
\rowcolor{Gray}
V~\cite{Vanhollebeke}BR~\cite{2013ApJ...779..115B} &  $202^{+8}_{-6}$  &  $16^{+1}_{-2}$ &  $9.4^{+1.2}_{-0.7}$ & $6.8^{+0.3}_{-0.4}$& $10.1 ^{+1.0}_{-0.7}$ & 0.20\\
BG~\cite{Bissantz:2001wx}BR~\cite{2013ApJ...779..115B} & $202^{+6}_{-8}$ &    $16^{+2}_{-1}$  & $9.5^{+0.8}_{-1.1}$ & $6.8^{+0.4}_{-0.4}$& $10.2 ^{+0.8}_{-0.9}$ &0.16\\
\rowcolor{Gray}
Z~\cite{Zhao:1995qh}BR~\cite{2013ApJ...779..115B} & $201^{+6}_{-7}$ &   $16^{+1}_{-2}$ & $9.3^{+0.9}_{-0.9}$ & $6.7^{+0.3}_{-0.4}$ & $10.0 ^{+0.7}_{-0.9}$ & 0.26 \\
R~\cite{Robin}BR~\cite{2013ApJ...779..115B} & $200^{+9}_{-5}$ &    $16^{+2}_{-1}$ &  $9.5^{+0.9}_{-1.0}$ & $6.8^{+0.3}_{-0.4}$& $10.1 ^{+0.8}_{-0.9}$ & 0.31 \\
\rowcolor{Gray}
G2~\cite{Stanek:1995ws}HG~\cite{2003ApJ...592..172H} & $193^{+8}_{-7}$ &   $19^{+2}_{-2}$  & $8.2^{+1.1}_{-0.9}$ &  $6.4^{+0.5}_{-0.4}$& $8.8 ^{+1.0}_{-0.7}$ & 0.56\\
E2~\cite{Stanek:1995ws}HG~\cite{2003ApJ...592..172H} & $193^{+7}_{-7}$ &  $19^{+2}_{-2}$ & $8.3^{+1.0}_{-0.9}$ & $6.4^{+0.4}_{-0.5}$& $8.8 ^{+1.0}_{-0.6}$ & 1.0\\
\rowcolor{Gray}
V~\cite{Vanhollebeke}HG~\cite{2003ApJ...592..172H} & $193^{+9}_{-7}$ & $19^{+3}_{-1}$ & $8.2^{+1.1}_{-0.9}$ & $6.4^{+0.4}_{-0.5}$& $8.9^{+0.9}_{-0.8}$ & 0.32\\
BG~\cite{Bissantz:2001wx}HG~\cite{2003ApJ...592..172H} & $192^{+8}_{-7}$ & $19^{+2}_{-2}$ & $8.1^{+1.0}_{-0.9}$ &$6.5^{+0.5}_{-0.4}$ & $8.8 ^{+0.9}_{-0.7}$&0.53\\
\rowcolor{Gray}
Z~\cite{Zhao:1995qh}HG~\cite{2003ApJ...592..172H}  & $193^{+9}_{-6}$ &  $19^{+2}_{-2}$ & $8.3^{+1.1}_{-0.9}$ & $6.4^{+0.4}_{-0.5}$& $8.9 ^{+1.0}_{-0.7}$& 0.75\\
R~\cite{Robin}HG~\cite{2003ApJ...592..172H} & $193^{+7}_{-7}$ & $19^{+2}_{-2}$ & $8.2^{+0.9}_{-0.9}$ &$6.3^{+0.4}_{-0.4}$ & $8.9 ^{+0.7}_{-0.8}$ & 1.43\\
\rowcolor{Gray}
G2~\cite{Stanek:1995ws}CM~\cite{2011MNRAS.416.1292C} & $188^{+11}_{-6}$ &  $22^{+2}_{-3}$  & $7.6^{+1.4}_{-0.7}$ & $6.2^{+0.5}_{-0.3}$& $8.4 ^{+1.0}_{-0.7}$& 0.09\\
E2~\cite{Stanek:1995ws}CM~\cite{2011MNRAS.416.1292C} & $186^{+8}_{-7}$ &  $22^{+2}_{-3}$ &  $7.5^{+1.3}_{-0.6}$ & $6.4^{+0.5}_{-0.4}$& $8.1 ^{+1.1}_{-0.5}$ & 0.19\\
\rowcolor{Gray}
V~\cite{Vanhollebeke}CM~\cite{2011MNRAS.416.1292C} & $191^{+8}_{-10}$ & $22^{+3}_{-2}$ & $7.9^{+1.0}_{-1.1}$ & $6.2^{+0.4}_{-0.5}$& $8.5 ^{+0.9}_{-1.0}$& 0.06 \\
BG~\cite{Bissantz:2001wx}CM~\cite{2011MNRAS.416.1292C} & $188^{+9}_{-6}$ &  $22^{+2}_{-3}$ &  $7.8^{+0.9}_{-1.0}$ & $6.3^{+0.5}_{-0.4}$& $8.4^{+0.7}_{-0.9}$& 0.09 \\
\rowcolor{Gray}
Z~\cite{Zhao:1995qh}CM~\cite{2011MNRAS.416.1292C}  & $189^{+8}_{-8}$  &  $22^{+3}_{-2}$ & $7.7^{+1.1}_{-0.9}$ & $6.2^{+0.5}_{-0.4}$& $8.5^{+0.9}_{-0.8}$ & 0.12\\
R~\cite{Robin}CM~\cite{2011MNRAS.416.1292C}  & $189^{+9}_{-6}$ &  $22^{+2}_{-2}$ & $7.9^{+1.0}_{-0.9}$ & $6.2^{+0.4}_{-0.4}$& $8.5^{+0.8}_{-0.9}$&0.25\\
\rowcolor{Gray}
G2~\cite{Stanek:1995ws}dJ~\cite{2010ApJ...714..663D} & $190^{+9}_{-8}$ &  $21^{+3}_{-2}$ & $8.0^{+1.1}_{-1.0}$ & $6.3^{+0.5}_{-0.3}$& $8.7 ^{+0.9}_{-0.9}$& 0.13\\
E2~\cite{Stanek:1995ws}dJ~\cite{2010ApJ...714..663D}  & $190^{+8}_{-7}$  &   $20^{+3}_{-2}$ & $7.9^{+1.0}_{-0.9}$ &  $6.5^{+0.5}_{-0.4}$& $8.5^{+0.9}_{-0.7}$&0.30\\
\rowcolor{Gray}
V~\cite{Vanhollebeke}dJ~\cite{2010ApJ...714..663D}  & $191^{+8}_{-9}$ & $21^{+3}_{-2}$& $8.0^{+1.4}_{-0.8}$ & $6.4^{+0.4}_{-0.5}$& $8.6 ^{+1.0}_{-0.8}$&0.08\\
BG~\cite{Bissantz:2001wx}dJ~\cite{2010ApJ...714..663D}   & $190^{+8}_{-8}$  & $21^{+3}_{-2}$ & $7.9^{+1.0}_{-1.0}$ & $6.4^{+0.5}_{-0.3}$& $8.7^{+0.7}_{-1.0}$&0.13\\
\rowcolor{Gray}
Z~\cite{Zhao:1995qh}dJ~\cite{2010ApJ...714..663D} & $188^{+11}_{-6}$ &  $21^{+3}_{-2}$ &  $7.8^{+1.1}_{-0.9}$ & $6.4^{+0.3}_{-0.4}$& $8.4^{+1.1}_{-0.7}$&0.17\\ 
R~\cite{Robin}dJ~\cite{2010ApJ...714..663D} & $189^{+9}_{-7}$ &  $21^{+2}_{-3}$ & $7.7^{+1.2}_{-0.8}$ & $6.4^{+0.4}_{-0.4}$& $8.4 ^{+1.0}_{-0.7}$&0.33\\
\rowcolor{Gray}
G2~\cite{Stanek:1995ws}J~\cite{Juric:2005zr} & $193^{+10}_{-7}$ &  $19^{+2}_{-2}$  & $8.2^{+1.3}_{-0.9}$ & $7.1^{+0.6}_{-0.4}$& $9.0 ^{+1.1}_{-0.8}$& 0.43\\
E2~\cite{Stanek:1995ws}J~\cite{Juric:2005zr} & $192^{+7}_{-9}$ & $19^{+2}_{-2}$  & $8.2^{+0.8}_{-1.2}$ & $7.3^{+0.4}_{-0.5}$& $9.0^{+0.7}_{-1.0}$& 0.73\\
\rowcolor{Gray}
V~\cite{Vanhollebeke}J~\cite{Juric:2005zr} & $192^{+10}_{-7}$ &  $20^{+3}_{-2}$ & $8.2^{+1.2}_{-1.0}$ & $7.1^{+0.6}_{-0.4}$& $8.9 ^{+1.1}_{-0.8}$&0.26\\
BG~\cite{Bissantz:2001wx}J~\cite{Juric:2005zr} & $192^{+9}_{-7}$ &  $20^{+2}_{-3}$ & $8.2^{+1.1}_{-1.0}$ & $7.2^{+0.5}_{-0.5}$& $8.9^{+1.0}_{-0.8}$&0.38\\
\rowcolor{Gray}
Z~\cite{Zhao:1995qh}J~\cite{Juric:2005zr} & $191^{+10}_{-7}$&  $19^{+3}_{-2}$ & $7.9^{+1.3}_{-0.8}$ & $7.1^{+0.5}_{-0.5}$& $8.8 ^{+0.9}_{-0.9}$&0.55\\
R~\cite{Robin}J~\cite{Juric:2005zr} & $191^{+10}_{-6}$ &  $20^{+2}_{-2}$ & $8.0^{+1.4}_{-0.7}$ & $7.1^{+0.5}_{-0.5}$& $8.8^{+1.0}_{-0.8}$&1.01\\
\hline
\end{tabular}}
\caption{
Maximum a posteriori (MAP) estimates with uncertainties obtained from the 68\% HPD region. All baryonic morphologies assume the gas density profile taken from \cite{0004-637X-497-2-759, Ferriere:2007yq}. The Bayes factor $B_{i0}$ is calculated using Eq.~\eqref{eqn:Bayes factor}.
$M_{\rm bar}$ corresponds to the total baryonic mass within $R_{200}$.}
\label{tab:virial_radius_mass} 
\end{table}

\section{Results}
\label{sec:results}

\begin{figure}
\centering
\includegraphics[width=0.9\textwidth]{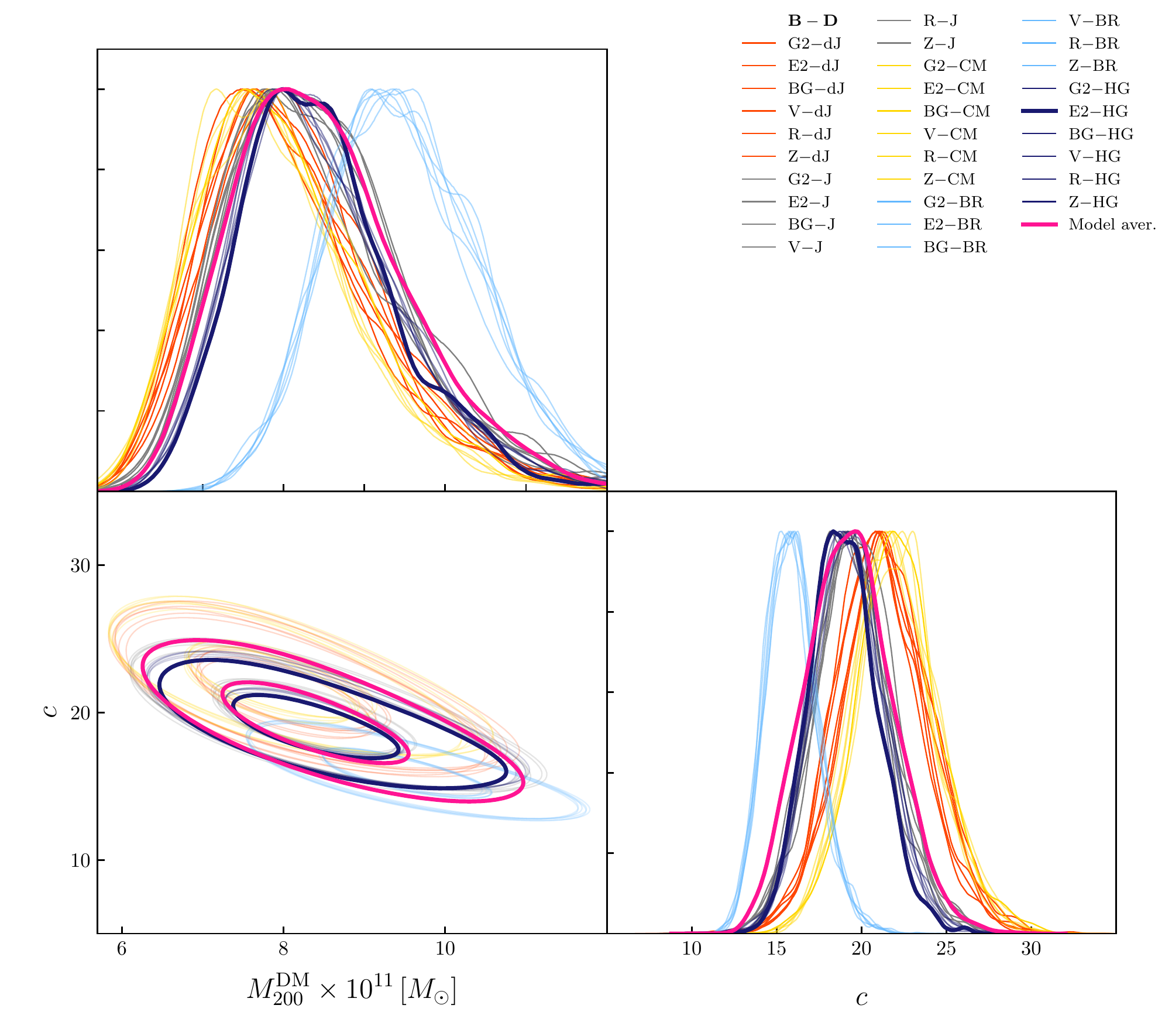}
\caption{\label{fig:posteriors_all_models} 
One and two-dimensional marginalized posterior distributions for the dark matter parameters $c$ and $M_{200}^{\rm DM}$ for our reference morphology (thick blue), all other baryonic morphologies (thin lines) and model-averaged (thick pink). In the legend caption, the first part of the name refers to the bulge morphology while the second part to the disk morphology (see table~\ref{tab:virial_radius_mass} for references). Lines are colored by disk morphology since the results are mainly dictated by the disk, which contains most of the baryonic mass, as noted in~\cite{2015JCAP...12..001P}.}
\end{figure}

\begin{figure}
\centering
\includegraphics[width=0.9\textwidth]{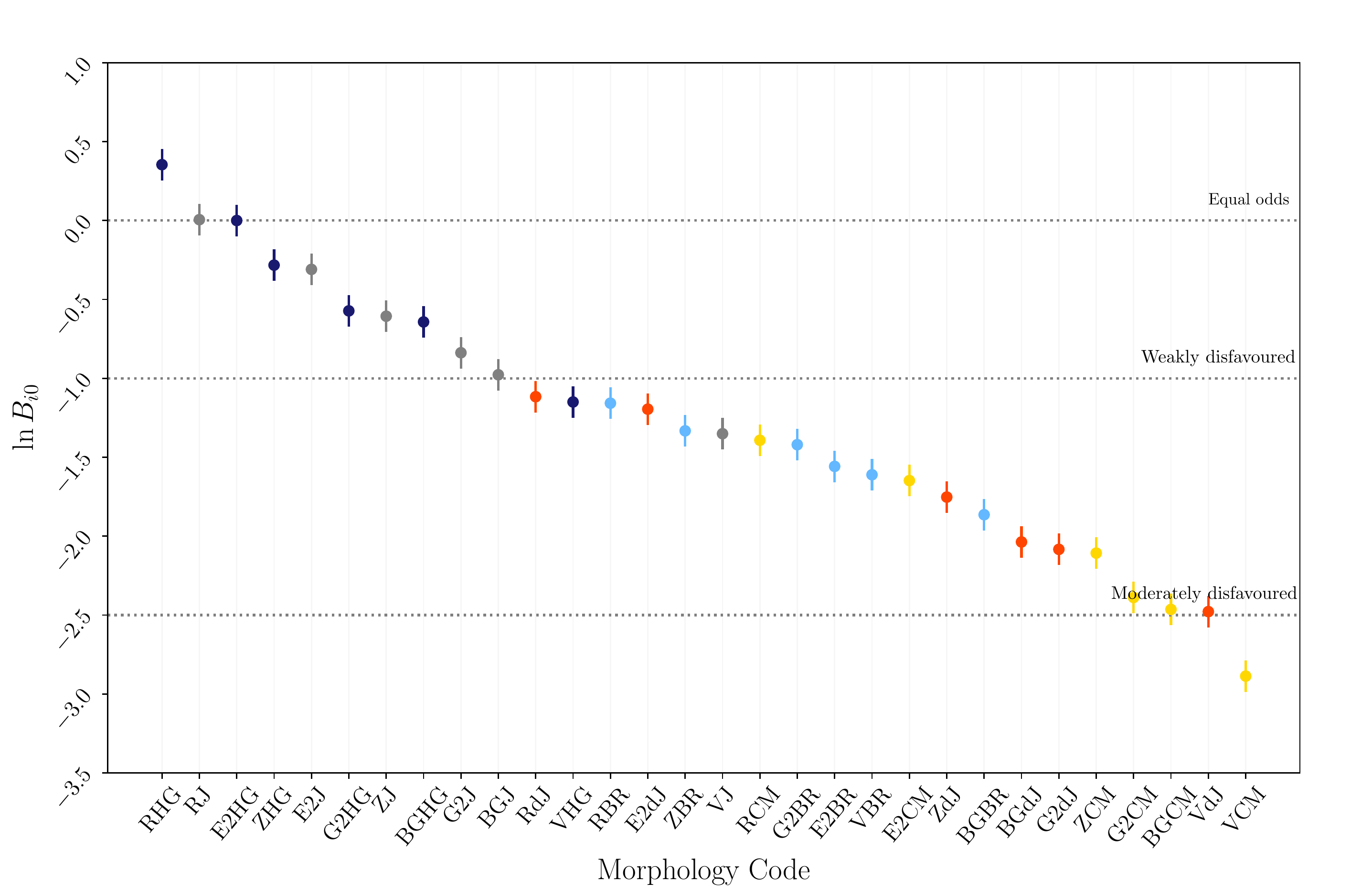}
\caption{\label{fig:BayesFactor_all_models} 
Natural logarithm of Bayes factors plotted for each baryonic morphology designation. Color-coding is the same as in figure~\ref{fig:posteriors_all_models}. Horizontal dotted lines delimit models that are weakly disfavoured ($\ln B_{i0} < -1.0$) and moderately disfavoured ($\ln B_{i0} < -2.5$) on the Jeffreys' scale of evidence as compared with the reference morphology E2HG.}
\end{figure}

In this section we present results obtained by using in a combined form the three sets of data  \texttt{galkin$_{12}$+Huang$_1$+Huang$_2$}. We show model-averaged parameter constraints, as well as constraints for individual morphologies. In Section~\ref{sec:tests} we show that these results are robust against several tests including the adoption of alternate halo priors and the use of different rotation curve data sets. Finally, on top of the uncertainties coming from our analysis we quantify two additional systematic uncertainties related to the parameterization of the underlying dark matter density profile and the value of the local circular velocity.

\subsection{Posterior constraints}

First, we present the results conditional on each of the 30 possible combinations of disk and bulge morphologies discussed in Section~\ref{sec:baryons}. table~\ref{tab:virial_radius_mass} summarizes the posterior constraints on the virial radius $R_{200}$, concentration parameter $c$, virial mass $M_{200}^{\rm DM}$, and baryonic mass of the Milky Way for each permuation of possible baryonic morphologies. It is interesting to note that both the virial mass and the baryonic mass do not vary much from one morphology to the next. Such small variations can also be appreciated in figure~\ref{fig:posteriors_all_models}, where we plot the resulting posteriors for the dark matter parameters $c$ and $M_{200}^{\rm DM}$ for different baryonic morphologies along with the model-averaged posterior described in Section~\ref{subsubsec:model_averaging}. 
We present the results only for the above mentioned two parameters because, as we showed in Paper~I, the slope of the inner dark matter density profile $\gamma$ and the scale radius $r_s$ are degenerate, thus making the separate reconstruction of the two parameters challenging. Here, with the conveniently-parameterized gNFW profile, we instead have a correlation between $\gamma$ and the concentration parameter $c$, with the former still remaining weakly constrained. Despite this degeneracy, the data yields tighter constraints on $M_{200}^{\rm DM}$, which is the primary target of this study.

We show in figure~\ref{fig:BayesFactor_all_models} the Bayes factors $\ln B_{i0}$ between all models and the reference morphology, together with levels that denote ``weak'' and ``moderate'' evidence against model $i$ (horizontal dotted lines), according to the nomenclature adopted by~\cite{Trotta_2008}. We find moderate evidence against only one morphology (VCM) when compared to the reference morphology, with all others having posterior odds of less than $12:1$. We also notice that most of the J (\cite{Juric:2005zr}) and HG (\cite{2003ApJ...592..172H}) disk types have Bayes factors above even the ``weak'' evidence threshold, meaning that they all contribute approximately equally to the model-averaged posterior\footnote{The morphology of the HG stellar disk \cite{2003ApJ...592..172H} is based on a pure thin plus thick disk, while the morphology of the J stellar disk \cite{Juric:2005zr} additionally includes a stellar halo component.}. No baryonic morphology can be ruled out with ``strong'' evidence, which would require $\ln B_{i0} = -5.0$, or odds in excess of $150:1$. This result is conditional on our choice of dark matter profile (described by a gNFW profile) and assumed value of the Sun's circular velocity. We address this point further in Section~\ref{sec:tests} below.

After marginalizing over all other parameters and model averaging over baryonic morphologies, we obtain the following determination of the Milky Way halo's dark matter virial mass $\log_{10}M_{200}^{\rm DM}/ {\rm M_\odot}~=~11.92^{+0.06}_{-0.05}$ or on a linear scale: 
\begin{equation}
M_{200}^{\rm DM}=8.3^{+1.2}_{-0.9}\times 10^{11}\,\mathrm{M}_{\odot},
\label{eqn:M200MA}
\end{equation} where uncertainties correspond to the 68\% credible region (defined as highest posterior density, HPD, interval, i.e., the shortest interval containing 68\% of posterior probability). Our estimate of the total mass of the Milky Way --the sum of baryons and dark matter-- within the virial radius, is $\log_{10}M_{\rm tot}/ {\rm M_\odot} = 11.95^{+0.04}_{-0.04}$ or on a linear scale:

\begin{equation}
M_{\rm tot}=8.9^{+1.0}_{-0.8}\times10^{11}\,\mathrm{M}_{\odot}.
\label{eqn:MtotMA}
\end{equation}
The quoted uncertainties on the above estimates take into account both statistical and systematic uncertainties, the latter arising due to our ignorance of the shape of the baryonic components in the Galaxy.

\section{Tests of robustness\label{sec:tests}}

\subsection{Average long-term properties of the MAP estimate}
\label{sec:accuracy}
Our analysis is Bayesian and all results are conditioned upon the actual data that was obtained. But it is informative to explore the frequentist performance of our method, in particular how it responds to expected fluctuations in the measurements.

To do this we generate 100 mock rotation curve data sets (with properties mimicking the real data) and perform our Bayesian analysis on each one. A mock observation is generated by fixing $\Phi = (c, M_{200}^{\rm DM}, \gamma, \Sigma_*, \left<\tau\right>)$ to a set of ``true values'' and calculating the resulting rotation curve $\omega_c(r,\Phi)$ as in Eq.~\eqref{eqn:vel_model}. The mock data for each radial bin $\bar{\omega}_i$ are sampled from a Gaussian with mean $\omega_c(r,\Phi)$ and standard deviation equal to the standard deviation of the real data within the bin $\sigma^2_{\bar{\omega},i}$, i.e. using the first factor of Eq.~\eqref{eqn:L}. This procedure for the generation of the mock data based on observational uncertainties is the same as adopted in Paper I. We fix the baryonic morphology to our reference morphology (E2HG) in both the mock data generation and the reconstruction.

We consider 25 fiducial configurations for $\Phi$ (5 possibilities each for $c$ and $\gamma$ and fixed values for the remaining three parameters). For each configuration we generate 100 mock rotation curve observations, construct the posterior for each using the identical settings as in our analysis above, and identify the MAP estimate of virial mass $\hat{M}_\mathrm{MAP}$. We quantify the performance of our procedure by estimating the fractional standard error, defined as $\mathrm{FSE}=\sqrt{\mathrm{E}[(\hat{M}_\mathrm{MAP}-M_\mathrm{true})^2]}/M_\mathrm{true}$, where $\mathrm{E}$ denotes expectation under repeated observations (for details see Paper~I, Section~4.1). We approximate this expectation by averaging over the 100 mock observations. 

We find that for all of our 25 fiducial configurations the $\mathrm{FSE}$ does not go above $\sim~20\%$. This is similar to the width of the marginalized posterior for $M_{200}^{\rm DM}$ conditioned both on the actual data and the mock data. From this we conclude that, first, the width of the posterior is comparable to that expected from random fluctuations in the data. This suggests that the posterior is likelihood-dominated. In other words, the posterior appears to be capturing the effects of measurement uncertainty as we might expect. Second, the priors are not inducing a significant frequentist bias in our analysis since the (fractional) bias in the $\hat{M}_\mathrm{MAP}$ estimator can be no larger than the FSE (see Section.~\ref{subsec:twosearches} for further analysis of prior dependence).

\begin{figure}
\centering
\includegraphics[width=0.9\textwidth]{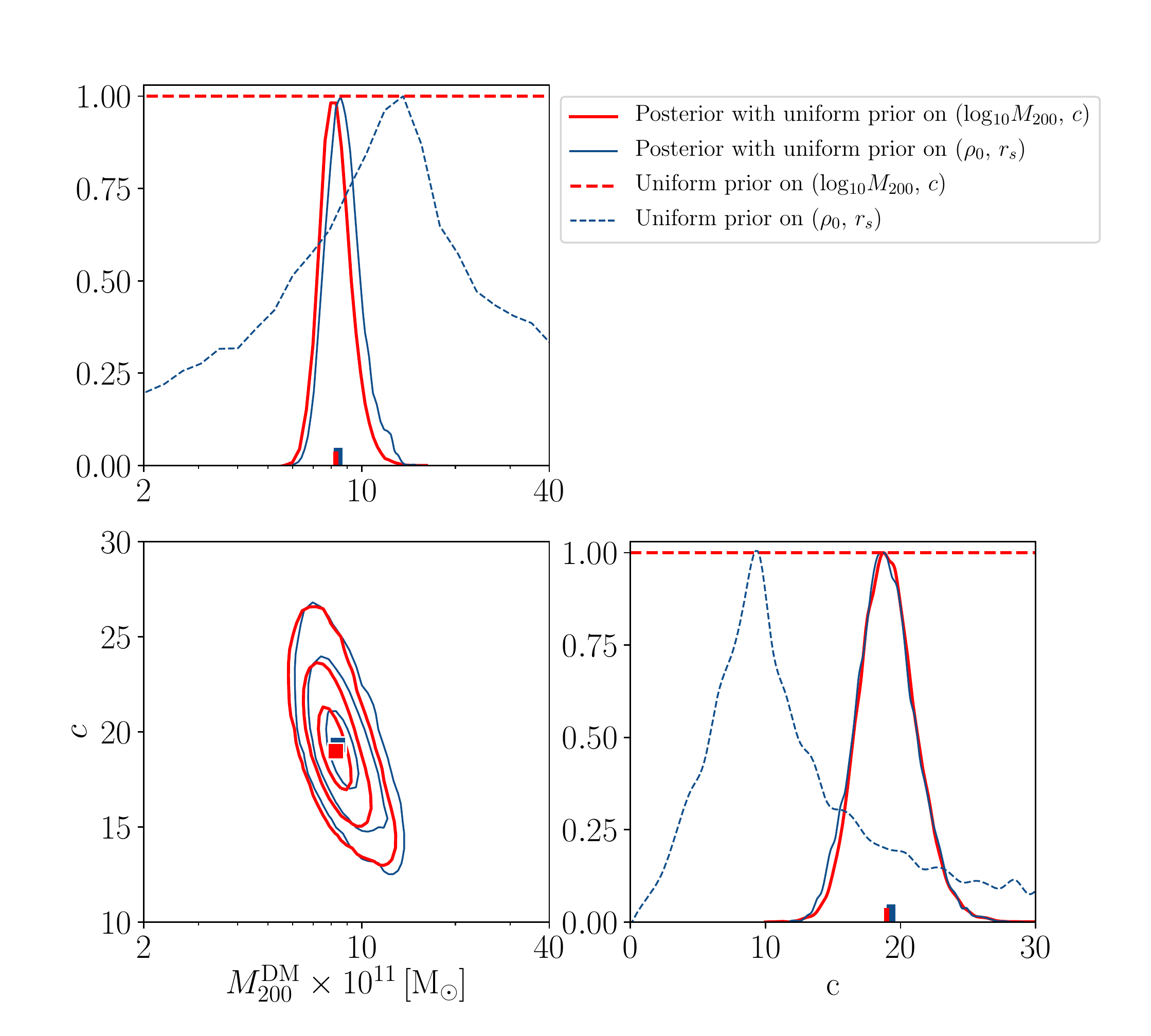}
\caption{\label{fig:corner_pl} 
One-dimensional marginal densities (normalized to the peak) and two-dimensional marginal posterior 68\%, 95\% and 99\% HPD regions with a uniform prior on $\log_{10}M_{200}^{\rm DM}$ and $\,c$ (red solid) and with a uniform prior on $\rho_0$ and $r_s$ (blue solid), assuming our reference baryonic morphology (E2HG). Dashed lines in the 1D plots indicate the corresponding priors. Squares/dashes give the maximum likelihood values in each case.}
\end{figure}

\subsection{Choice of priors}
\label{subsec:twosearches}
\label{App:twoapproach}

The results described above have been obtained by adopting the set $(c, \log_{10}M_{200}^{\rm DM}, \gamma)$ as parameters for the dark matter halo,  with  uniform priors described in Eq.~\eqref{eqn:priors}. As in any Bayesian analysis the choice of prior distribution is ultimately subjective and it is important to quantify how results depend on this choice.

We consider the alternative parameterization of the gNFW profile in terms of $\gamma$, $\rho_0$, and $r_s$ (see Eq.~\eqref{eqn:gNFW}) and consider uniform priors on these parameters as in Paper~I. The range allowed for each parameter is as follows:
\begin{equation}\label{eqn:priors_g_rs_rho0}
\begin{aligned}
    \gamma &\in [0, 3],\\
    \frac{r_s}{\mathrm{kpc}} &\in [0,40],\\
    \frac{\rho_0}{\mathrm{GeV/kpc^3}} &\in [0,1].
\end{aligned}
\end{equation}

It is to be noted that the upper edge of the prior range of $\gamma$ is larger than the one from Eq.~\eqref{eqn:priors}. However, our results are insensitive to this choice since the posterior always constrains $\gamma$ to be less than 2 (cf.~figure~8 of Paper I).

Since the relationship between the two parameterizations is non-linear, uniform priors in $(r_s, \rho_0)$ do not correspond to uniform priors in $(c, M_{200}^{\rm DM})$. However, if the likelihood is sufficiently constraining (i.e. data-dominated) we expect the two posterior distributions to agree. 
This is demonstrated in figure~\ref{fig:corner_pl}, where we compare the posterior distributions on $(\log_{10}M_{200}^{\rm DM},\,c)$ obtained with the two sets of priors and conditioned on the reference baryonic morphology E2HG. We observe that while a uniform prior on $(r_s,\,\rho_0)$ translates into an informative prior on $(c,\,M_{200}^{\rm DM})$, the posterior distributions obtained with the two sets of priors closely agree with each other (compare the red and blue solid curves). We thus conclude that the choice of parameterization has very little influence on our determinations of $c$ and $M_{200}^{\rm DM}$.

\begin{table}
\centering
\addtolength{\tabcolsep}{6pt}
\scalebox{1}{%
\begin{tabular}{ | l | c | c | c |}
\cline{2-3}
\multicolumn{1}{c|}{}& \multicolumn{2}{c|}{Uniform prior in} & \multicolumn{1}{c}{} \\
\multicolumn{1}{c|}{}& \multicolumn{1}{c}{$(\gamma,\log_{10}M_{200}^{\rm DM},c)$} & \multicolumn{1}{c|}{$(\gamma, r_s, \rho_0)$} & \multicolumn{1}{c}{} \\
\hline
Baryonic morphology & $\ln B_{i0}^{M_{200}^{\rm DM},c}$ & $\ln B_{i0}^{r_s,\rho_0}$ & $\ln B_{i0}^{M_{200}^{\rm DM},c} - \ln B_{i0}^{r_s,\rho_0}$\\
\hline
G2CM &  $-2.39  \pm0.10 $ &  $-2.37  \pm0.10 $ & $-0.02  \pm0.14$\\
VBR  &  $-1.61 \pm0.11 $ &  $-1.73 \pm0.10 $ & $0.12  \pm0.15$\\
BGHG &  $-0.64 \pm0.10 $ &  $-0.54 \pm0.10 $ & $-0.11  \pm0.14$\\
RJ   &  $0.01  \pm0.10 $ &  $0.04  \pm0.10 $ & $-0.03  \pm0.14$\\

\hline
\end{tabular}}
\caption{The Bayes factors $B_{i0}$ for two different choices of prior for the dark matter halo parameters for several baryonic morphologies. The final column shows the difference in Bayes factors between the two priors.}
\label{tab:bayesfactors_priors} 
\end{table}

To check the effect of prior choice on the Bayesian model averaging we compute the Bayes factors $B_{i0}$ (Section~\ref{subsubsec:model_averaging}) for several baryonic morphologies for the two prior choices. table~\ref{tab:bayesfactors_priors} shows that the changes in Bayes factors are negligible and so the weighting of each morphology in the model averaging is approximately independent of prior choice. The exercise indicates that our results are dominated by the observational data and are robust to changes to our prior distributions.

\subsection{Data selection}
\label{sec:data sets combination}

Our analysis combines data sets that are based on a variety of kinematic tracers (which either belong to the stellar disk or stellar halo). In this section we check how our results change when adopting different data combinations. Here we show the comparison of virial mass posteriors when analyzing the following combinations of data sets: 
\begin{enumerate}
\item \texttt{galkin$_{12}$} alone 
\item Huang$_1$ + Huang$_{2}$ (i.e. the full Huang~{\em et~al.}~\cite{2016MNRAS.463.2623H} data set over the range 8 to 100~kpc) 
\item \texttt{galkin$_{12}$} + Huang$_{1}$ + Huang$_{2}$ (the set used for our main analysis)
\end{enumerate}
For the sake of simplicity we fix the baryonic morphology to our reference model E2HG.

\begin{figure}
\centering
\includegraphics[width=0.9\textwidth]{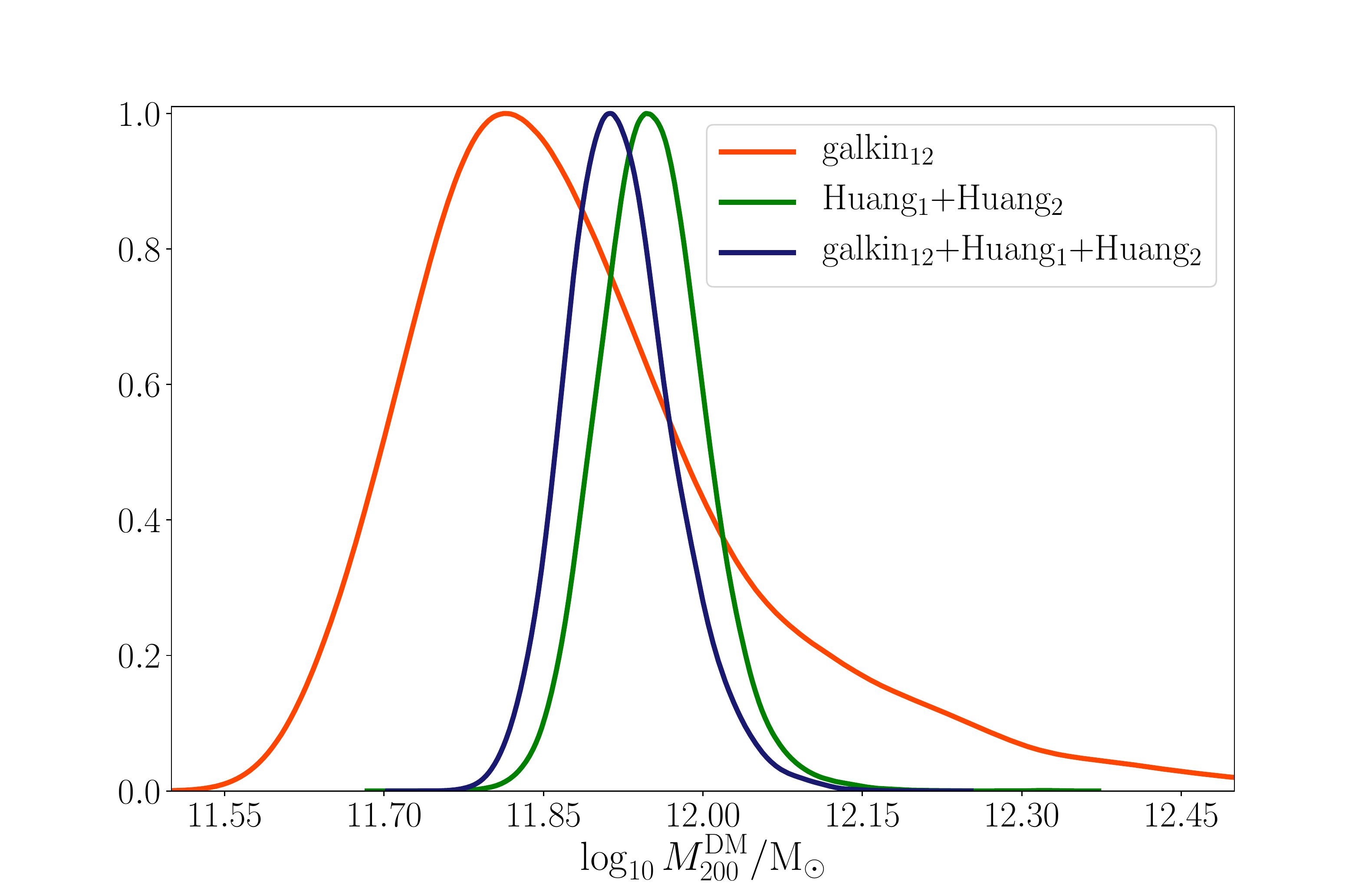}
\caption{\label{fig:virial mass different data sets} 
One-dimensional marginalized posterior distributions for the virial mass for our reference baryonic morphology (E2HG) when analyzing different data set combinations. The combination \texttt{galkin$_{12}$}+Huang$_1$+Huang$_2$ is the one used in our main analysis.}
\end{figure}

In figure~\ref{fig:virial mass different data sets} we show the posterior distributions on the virial mass for the three different combinations of data sets (numerical values are listed in table~\ref{tab:virial_mass_datasets}). The constraints on the virial mass from different data sets combinations are mutually compatible within the quoted statistical uncertainties. The difference between central MAP values for {\tt galkin}$_{12}$ and Huang$_1$ + Huang$_2$ data combinations (two data sets that are statistically independent) is $\Delta\log_{10}M_{200}^{\rm DM}=0.12\,\rm dex$. The posterior for the combined data set lies between that obtained from each data set separately, as expected.

\begin{table}
\centering
\addtolength{\tabcolsep}{6pt}
\scalebox{1}{%
\begin{tabular}{ | l | c | c |}
\hline 
Data combination & $\log_{10}M_{200}^{\rm DM}/\rm M_{\odot}$ & $\log_{10}M_{200}^{\rm DM}/\rm M_{\odot}$ \\
  & MAP & median  \\
\hline
\rowcolor{Gray}
\texttt{galkin}$_{12}$ & $11.83^{+0.16}_{-0.13}$ & $11.86^{+0.19}_{-0.12}$ \\
\texttt{Huang}$_{1}$ + Huang$_{1}$ & $11.95^{+0.05}_{-0.05}$ & $11.95^{+0.05}_{-0.05}$\\
\rowcolor{Gray}
 \texttt{galkin$_{12}$+Huang$_1$+Huang$_2$} & $11.92^{+0.05}_{-0.05}$ & $11.92^{+0.05}_{-0.05}$\\
\hline
\end{tabular}}
\caption{Estimate of dark matter virial mass for analyses of different data set combinations for our reference baryonic morphology (E2HG). The second column lists maximum a posteriori (MAP) estimates with uncertainties obtained from the 68\% HPD region. The third column gives the median of the posterior with uncertainties corresponding to the 15.9 and 84.1 percentiles of the posterior.}
\label{tab:virial_mass_datasets} 
\end{table}

\subsection{Choice of dark matter density profile} \label{sec:density profiles}

We examine the robustness of the Milky Way mass estimate with respect to the choice of dark matter density profile. In addition to the gNFW profile adopted above, we present here a comparison to the Einasto~\cite{1965TrAlm...5...87E} and Burkert profiles~\cite{1995ApJ...447L..25B} introduced in Eqs.~\eqref{eqn:Einasto} and~\eqref{eqn:Burkert}. 
For these two profiles we adopt uniform priors over the following ranges:

Einasto profile:
\begin{equation}\label{eqn:priors_Einasto}
\begin{aligned}
    c &\in [0, 50],\\
    \log_{10}\frac{M_{200}^{\rm DM}}{\mathrm{M}_\odot} &\in [10,13],\\
    \alpha &\in [0.1,1.5].
\end{aligned}
\end{equation}

Burkert profile:

\begin{equation}\label{eqn:priors_Burkert}
\begin{aligned}
    \frac{r_c}{\mathrm{kpc}} &\in [0,12],\\
    \log_{10}\frac{M_{200}^{\rm DM}}{\mathrm{M}_\odot} &\in [10,13].
\end{aligned}
\end{equation}

\begin{table}
\centering
\addtolength{\tabcolsep}{6pt}
\scalebox{1}{%
\begin{tabular}{ | l | c | c | c |}
\hline 
Profile & $\log_{10}M_{200}^{\rm DM}/\rm M_{\odot}$ & $\log_{10}M_{\rm bar}/\rm M_{\odot}$ & $\log_{10}M_{\rm tot}/\rm M_{\odot}$ \\

\hline
\rowcolor{Gray}
Einasto & $11.64^{+0.09}_{-0.07}$ & $ 10.88^{+0.03}_{-0.02}$ & $11.70^{+0.07}_{-0.06}$ \\
Burkert & $11.90^{+0.04}_{-0.03}$ & $10.88^{+0.02}_{-0.02}$ & $11.94^{+0.03}_{-0.03}$ \\
\rowcolor{Gray}
gNFW &  $11.92^{+0.06}_{-0.05}$ & $10.82^{+0.04}_{-0.03}$ & $11.95^{+0.05}_{-0.04}$ \\
\hline
\end{tabular}}
\caption{Virial, baryonic, and total mass estimates for three different dark matter density profiles. In each case, values are obtained after model averaging over baryonic morphologies. Central values are maximum a posteriori (MAP) estimates and uncertainties correspond to 68\% highest posterior density (HPD) credible intervals.
\label{tab:mass_dmprofiles}}
\end{table}

The resulting posterior estimates of the dark matter, baryonic, and total mass are given in table~\ref{tab:mass_dmprofiles}, obtained after marginalizing over all the other parameters and model averaging over baryonic morphologies.

\begin{figure}
\centering
\includegraphics[width=0.7\textwidth]{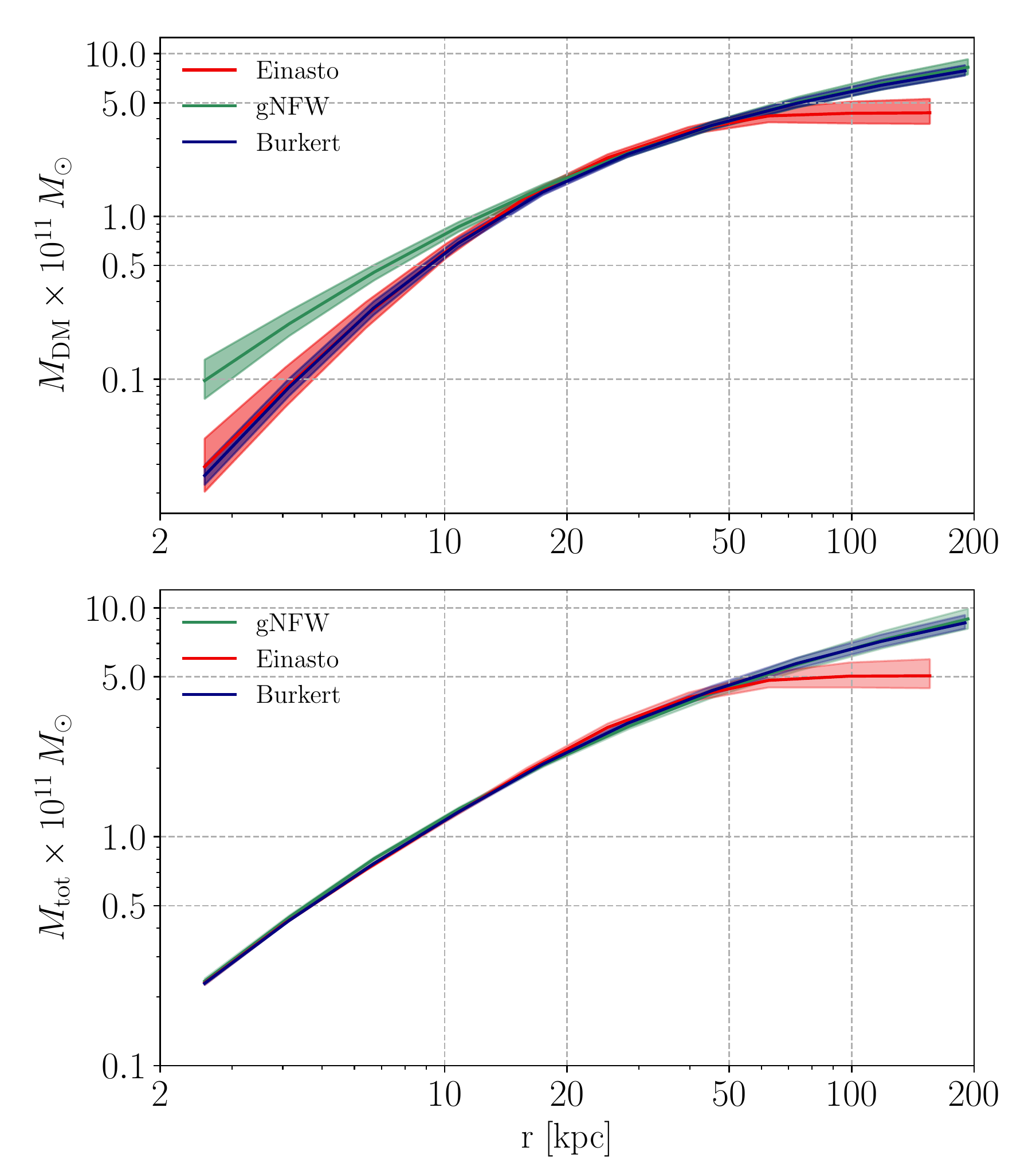}
\caption{\label{fig:mass_profiles_dif_par} 
MAP estimate of the cumulative mass profile (dark matter only in the top panel, total mass in the bottom panel) for the three different dark matter profiles (solid lines) and the corresponding 68\% credible intervals (light shaded areas). Credible intervals are HPD regions, conditioned on the radius and model-averaged over baryonic morphologies. Mass profiles are plotted out to the MAP estimate of the virial radius, $R_{200}$, for each profile.}
\end{figure}

The MAP estimates for $M_{200}^{\rm DM}$ obtained assuming gNFW and Burkert profiles are within the 68\% credible intervals of one another, but are both considerably larger than the value obtained assuming an Einasto profile. This can be understood from figure~\ref{fig:mass_profiles_dif_par}, showing the enclosed dark matter (top panel) and total mass (bottom panel) as a function of radius for the three profile types. The effect of profile choice on total mass is subdominant with respect to statistical uncertainties within a radius of about~50~kpc. However, the dark matter mass is determined independently of the assumed profile shape only in the range between 20 and 50~kpc (top panel in figure~\ref{fig:mass_profiles_dif_par}). Because the Einasto profile ties together the behavior of the inner and outer halo, the data-driven preference for a somewhat more cored dark matter profile in the inner 10~kpc (compared to gNFW) translates into a flatter cumulative mass profile beyond about 50~kpc, the region where approximately 50\% of the total mass is accumulated in the gNFW and Burkert cases. This explains why the Einasto profile gives a MAP estimate 0.28 (0.25) dex lower for $\log_{10}M_{200}^{\rm DM}$ ($\log_{10}M_{\rm tot}$) compared to gNFW or Burkert profiles. 

\begin{figure}
\centering
\includegraphics[width=0.7\textwidth]{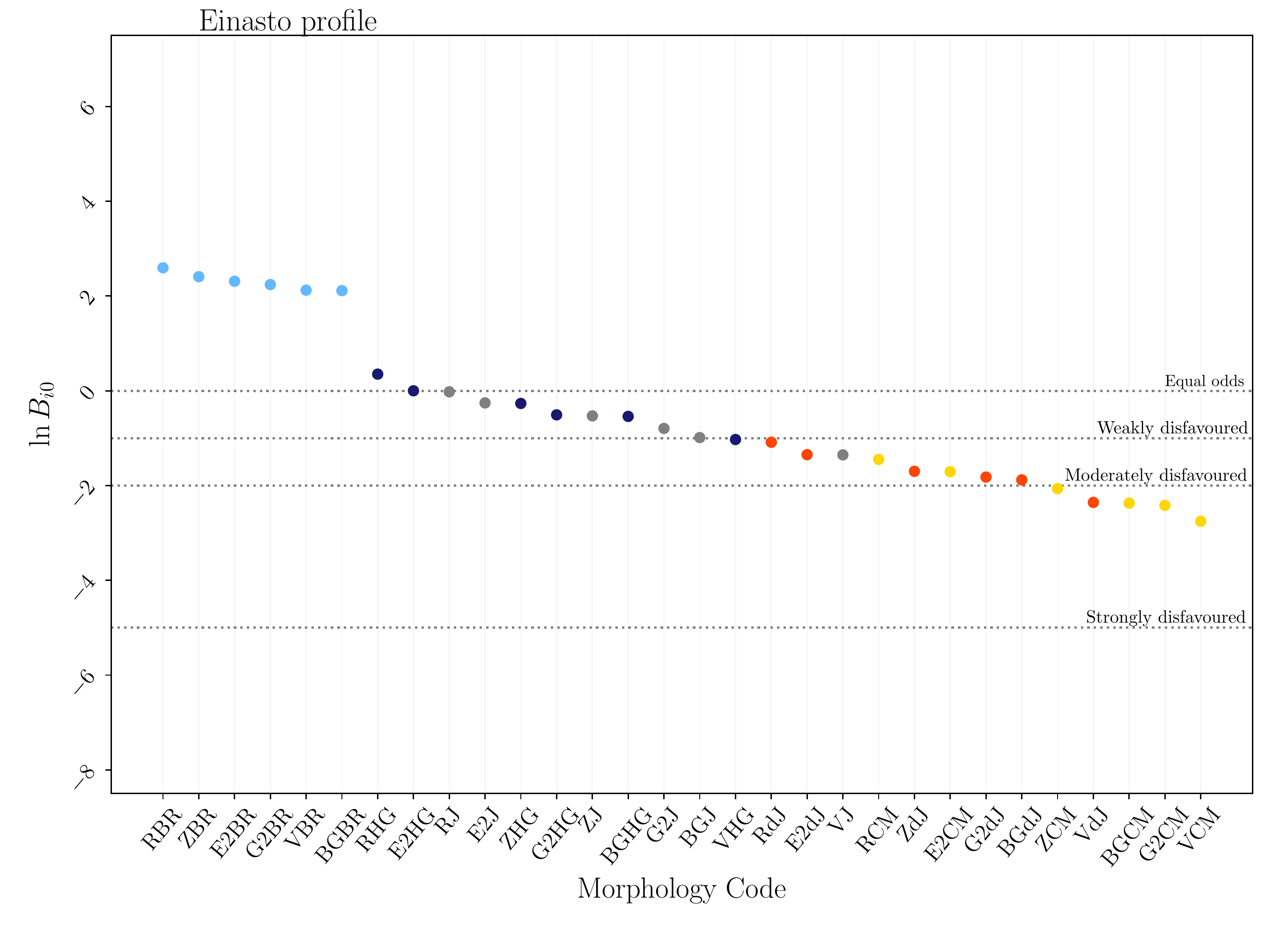} \\ 
\includegraphics[width=0.7\textwidth]{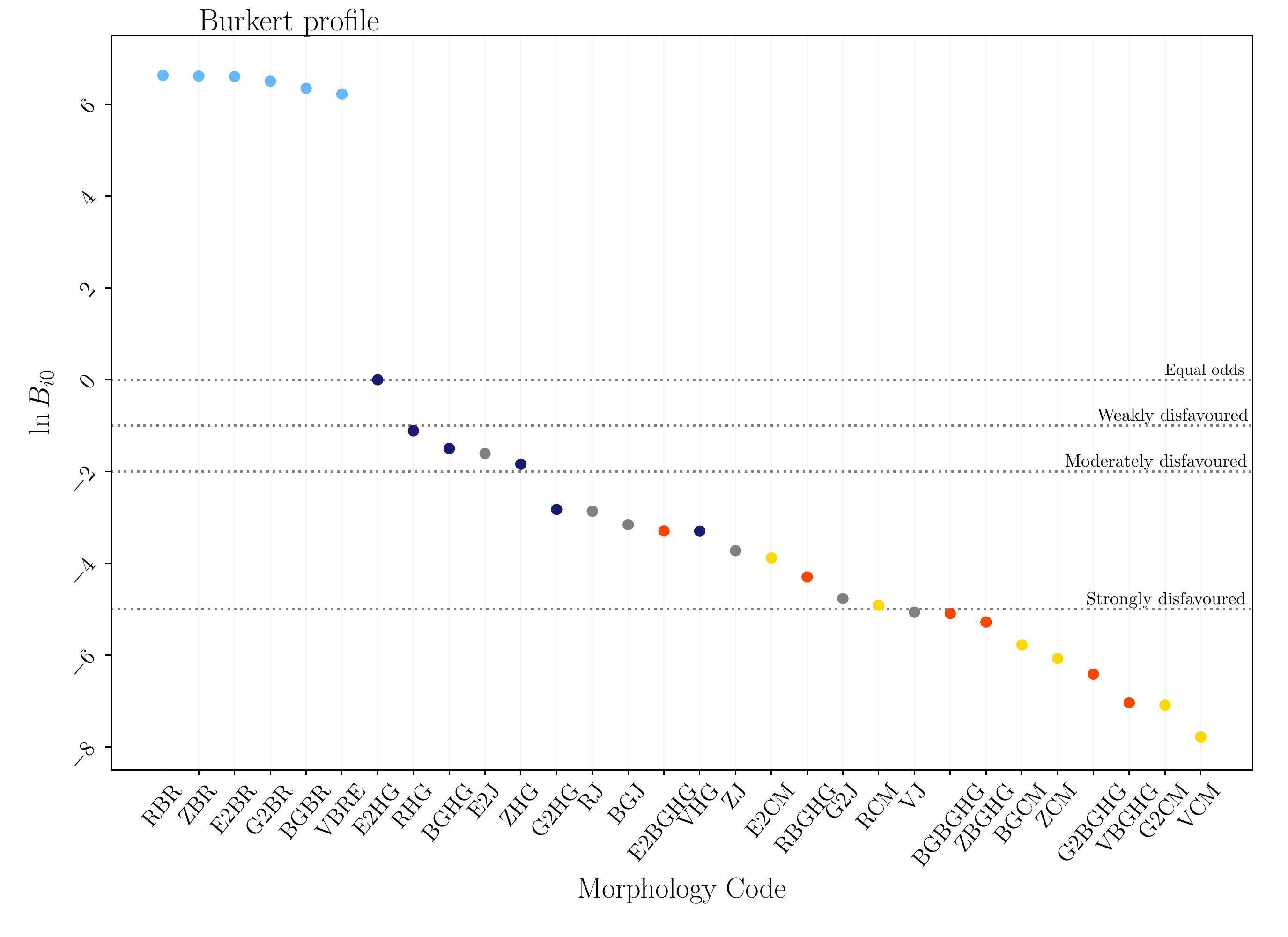}
\caption{\label{fig:Bayes_Factors_2} 
As in figure~\ref{fig:BayesFactor_all_models}, but for the Einasto (top panel) and Burkert (bottom panel) profiles.  Uncertanties on the Bayes factor are about $\ln \Delta B_{i0}=0.1$, smaller than the size of the marker on this scale.}
\end{figure}

In principle, a more flexible gNFW model could be adopted (e.g. with variable outer slope and variable sharpness of the transition around $r_s$). In such a model the width of the marginal posterior on $M_{200}^{\rm DM}$ would incorporate the additional systematic uncertainty due to the functional form of the dark matter density profile adopted. 
For the purposes of this study, we separate out the effect of profile choice from other sources of uncertainty, identifying it as source of systematic uncertainty for the virial mass $M_{200}^{\rm DM}$. We quantify such uncertainty by the difference in the MAP estimate for $\log_{10}M_{200}^{\rm DM}$ between the gNFW and the Einasto profile (since the Burkert profile gives a similar value as gNFW), and evaluate it to be 0.28~dex. 

The choice of dark matter profile also impacts on the model-averaging results, by changing the relative weights of the baryonic morphologies, an effect that feeds into (and is already accounted for by) the above systematic uncertainty. We have re-computed all Bayes factors entering into Eq.~\eqref{eqn:BMA2} for the Einasto and Burkert profiles, and they are plotted in figure~\ref{fig:Bayes_Factors_2}. Compared to figure~\ref{fig:BayesFactor_all_models}, we observe a preference for the BR-type disks, moderate in the case of the Einasto profile and strong for Burkert. This results in the model-averaged posteriors for these two profiles being strongly dominated by BR-type morphologies, differently from the gNFW case, where no morphology is strongly preferred. 

The preference for BR-type morphology for the Einasto and Burkert dark matter profile choice arises from a combination of two factors: firstly, BR-type morphologies allow for a better fit to the microlensing optical depth, $\langle \tau \rangle$; secondly, the BR-type morphologies exhibit a reduced Occam's razor effect in comparison to the other choices of morphologies. The latter is a purely Bayesian effect in our model comparison framework, arising from the different volume of the posterior distribution for the different morphologies when assuming one or the other dark matter profile. 

Indeed, we have checked that from a frequentist point of view, the preference for BR-type morphologies is weaker than in the Bayesian case. This is rather unusual, for in the more commonly encountered case of nested models the contrary is typically true: the Bayesian model comparison result is more conservative than hypothesis testing based on e.g. likelihood ratio tests. However, in this case the models being compared (i.e., different morphologies for the same choice of underlying dark matter profile) are not nested, so we cannot rely on the usual theorems regarding the distribution of the likelihood ratio test statistics. As an illustration, we have computed the distribution of the log-maximum likelihood ratio between two morphologies, BGBR and E2HG (our reference morphology) numerically, by producing an empirical distribution from mock data under each hypothesis. Inspection of the simulated distributions and comparison with the observed values of the test statistics show that, firstly, neither morphology can be rejected in a frequentist hypothesis test at any confidence level. This means that either morphology can be adequately fit with the data. Secondly, comparing morphologies via a  log-likelihood ratio test between the two hypotheses as test statistics results at best in a weak preference for one of the two ($p$-value of 0.02).

\subsection{Dependence on the local circular velocity}
\label{subsec:V0}

\begin{table}
\label{tab:V0_mass}
\centering
\addtolength{\tabcolsep}{6pt}
\scalebox{1}{%
\begin{tabular}{ | c | c |}
\hline 
$V_0$ [km/s] & $\log_{10}M_{200}^{\rm DM}/\rm M_{\odot}$ \\
\hline 
218       & $11.43^{+0.15\,(0.43)}_{-0.13\,(0.22)}$  \\[1ex] 
233       & $11.73^{+0.16\,(0.50)}_{-0.15\,(0.24)}$  \\[1ex]
239.89 (fiducial)  & $11.83^{+0.16\,(0.45)}_{-0.13\,(0.22)}$ \\[1ex]
248       & $11.97^{+0.17\,(0.53)}_{-0.14\,(0.24)}$  \\[1ex] 
\hline
\end{tabular}}
\caption{MAP values of the virial mass and the corresponding 68\%\,(95\%) credible intervals. We adopt our reference morphology, $R_0=8.34\,\rm kpc$ and use the {\tt galkin}$_{12}$ rotation curve data only, which allow for rescaling of $V_0$.}
\end{table}

\begin{figure}
\centering
\includegraphics[width=0.60\textwidth]{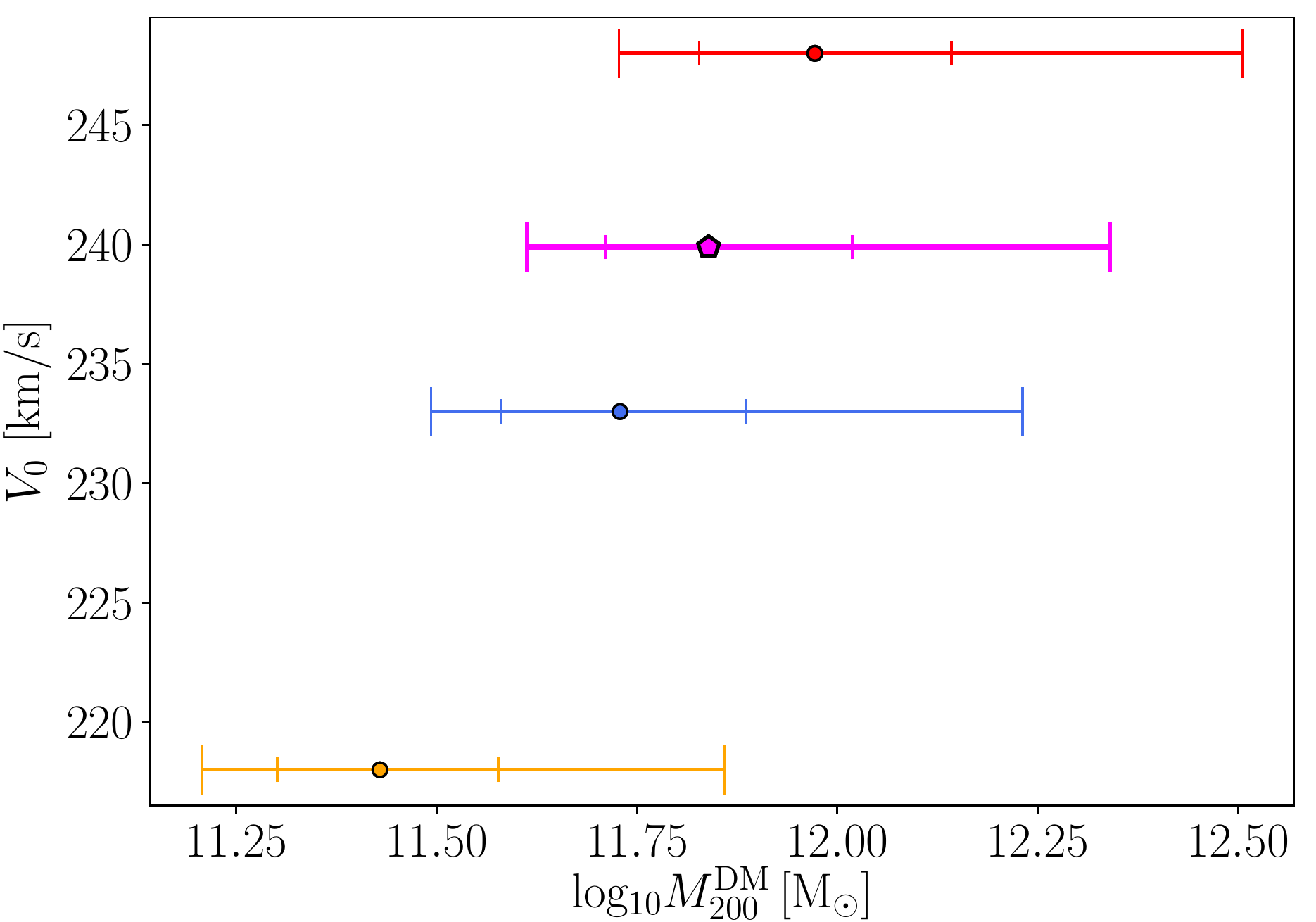}
\caption{\label{fig:V0_mass} Virial mass estimate as a function of local circular velocity $V_0$, for $R_0=8.34\,\rm kpc$.
The magenta pentagon shows the MAP estimate of the virial mass for the fiducial value of the circular velocity, $V_0=239.89\, \rm km/s$ see Section~\ref{sec:data sets}. Error bars are 68\%/95\% HPD credible regions. These constraints use the {\tt galkin}$_{12}$ rotation curve data only, which allow for rescaling of $V_0$, and assume our reference baryonic morphology (E2HG), without model-averaging over morphologies. The lower and upper values of $V_0$ enclose the 1-$\sigma$ region for $V_0$ from the latest measurements of the galactic parameters.}
\end{figure}

Rotation curve circular velocities --and ultimately our results-- depend on the galactic parameters $(R_0, V_0)$ adopted, and it is therefore important to test the solidity of our determination with respect to their variation. 
We note that the effect of varying $V_0$ dominates that of $R_0$, so in the following we focus on the effect $V_0$ has on the determination of the virial mass\footnote{It is well known that the variation of $R_0$ alone has smaller effects than that of $V_0$, within comparable ranges.
We have checked that for the case at hand, the variation of $R_0$ alone has negligible effects on estimate of the virial mass with respect to the variation of $V_0$, within the uncertaity intervals adopted \cite{Abuter:2018drb, BlandHawthorn&Ortwin2016}.}.

\par In order to perform such a test, we  make use of the \texttt{galkin$_{12}$} data alone: whereas it is trivial to rescale appropriately the rotation curve data from disk tracer measurements (such as those collected in \texttt{galkin}, see e.~g.~equations 1 and 2 in \cite{Pato:2017yai}) for different galactic parameters, it would be  extremely cumbersome to do the same for the  Huang~{\it et~al.}~\cite{2016MNRAS.463.2623H} dataset, as the halo objects are connected to an equivalent circular velocity in the disk through a full Jeans analysis.  In Section~\ref{sec:data sets combination} we showed that the determinations of $M_{200}^{\rm DM}$ are consistent when using either \texttt{galkin$_{12}$} alone, Huang$_1$+Huang$_2$, or both together. Therefore, we can explore the effect on $M_{200}^{\rm DM}$ of varying the galactic parameters using the \texttt{galkin$_{12}$} data set by itself, which allows for a simple rescaling of the adopted value of $V_0$.

We have so far adopted the galactic parameters in Huang~{\em et~al.}~\cite{2016MNRAS.463.2623H}, namely $R_0=8.34\,\mathrm{kpc}$ and $V_0=239.89\,\mathrm{km/s}$.
Here we vary $V_0$, highlighting that a wide range in $V_0$ encompasses uncertainties on the tangential peculiar motion of the Sun since the galactic parameters $R_0$, $V_0$ and $V_{\odot}$ are related through the total angular velocity of the Sun $\Omega_{g, \odot}$ \cite{BlandHawthorn&Ortwin2016, Benito+2019}.
Figure~\ref{fig:V0_mass} shows how the posterior on the virial mass (conditioned on the reference baryonic morphology E2HG)  is affected by changes in $V_0$. Numerical values are listed in table~\ref{tab:V0_mass} where we see that $\log_{10}M_{200}^{\rm DM}$ increases by a factor 0.54~dex  (or by a factor $\sim 3.5$) when varying $V_0$ from $218\,\rm km/s$ to $248\,\rm km/s$.
This range of values --- broader than the one in \cite{2018arXiv181011468E} $V_0= 233\pm 3$ km/s --- is based on the following where we propagate a set of astrophysical uncertainties.

We adopt the recent determinations of the galactic parameters ($R_0=8.122 \pm 0.031\,\mathrm{kpc}$ \cite{Abuter:2018drb}, the peculiar motion of the Sun in the tangential direction $V_{\odot}=12.24 \pm 0.47\,\mathrm{km/s}$ \cite{Schooenrich+2010}; $\Omega_{g, \odot} = 30.24\pm 0.12$ km/s/kpc \cite{ReidBrunthaler2004}, and the local standard of rest $V_{LSR}=0\pm15\,\mathrm{km/s}$ \cite{BlandHawthorn&Ortwin2016}) and use standard error propagation to obtain a $V_0$ distribution described by the above-mentioned range, $V_0= 233\pm 15$ km/s. Notice that the value $V_0=239.89\,{\rm km/s}$ adopted as fiducial throughout this paper, is within this interval, not too far off with respect to the best current estimate. If we vary the value of $V_0$ within the $1\sigma$ interval (i.e., from $218$ km/s to $248$ km/s), our MAP estimate of $\log_{10}M_{200}^{\rm DM}$ (obtained from \texttt{galkin}$_{12}$ data only) varies by 0.42~dex for the Burkert profile, by 0.54~dex for the gNFW profile, and by 0.54~dex for the Einasto profile (assuming the reference morphology in all cases; for Einasto and Burkert profiles and BR morphology the variation is about 0.40~dex). In order to be conservative, we thus adopt the largest of these variations, namely 0.54~dex. Consequently, our estimate of the systematic uncertainty associated with the residual uncertainty in the value of $V_0$ is half of this value: 0.27~dex (a factor of 1.9 on a linear scale). This is in addition to the systematic uncertainty due to the choice of dark matter profile, which is comparable at 0.28~dex. By the same procedure we estimate the systematic uncertainty in $\log_{10}M_\mathrm{tot}$ due to $V_0$ and find it to be similar at 0.25~dex (a factor of 1.8 on a linear scale).

Finally, a change in the value of $V_0$ adopted also induces a change in the Bayes factors for the baryonic morphologies, and hence an additional change in the inferred value of the mass as the weight of each morphology shifts. While this effect is not captured by our estimate above for the systematic uncertainty from $V_0$, it could in the future be addressed by upgrading $V_0$ to a nuisance parameter to be included in the scan. This will however require addressing the issue of how to perform an on-the-fly Jeans analysis (as the parameters in the model are scanned over) in order to obtain a $V_0$-dependent likelihood for the Huang~{\em et~al.} data, something that we leave for future work.

\begin{figure}
\centering
\includegraphics[width=0.95\textwidth]{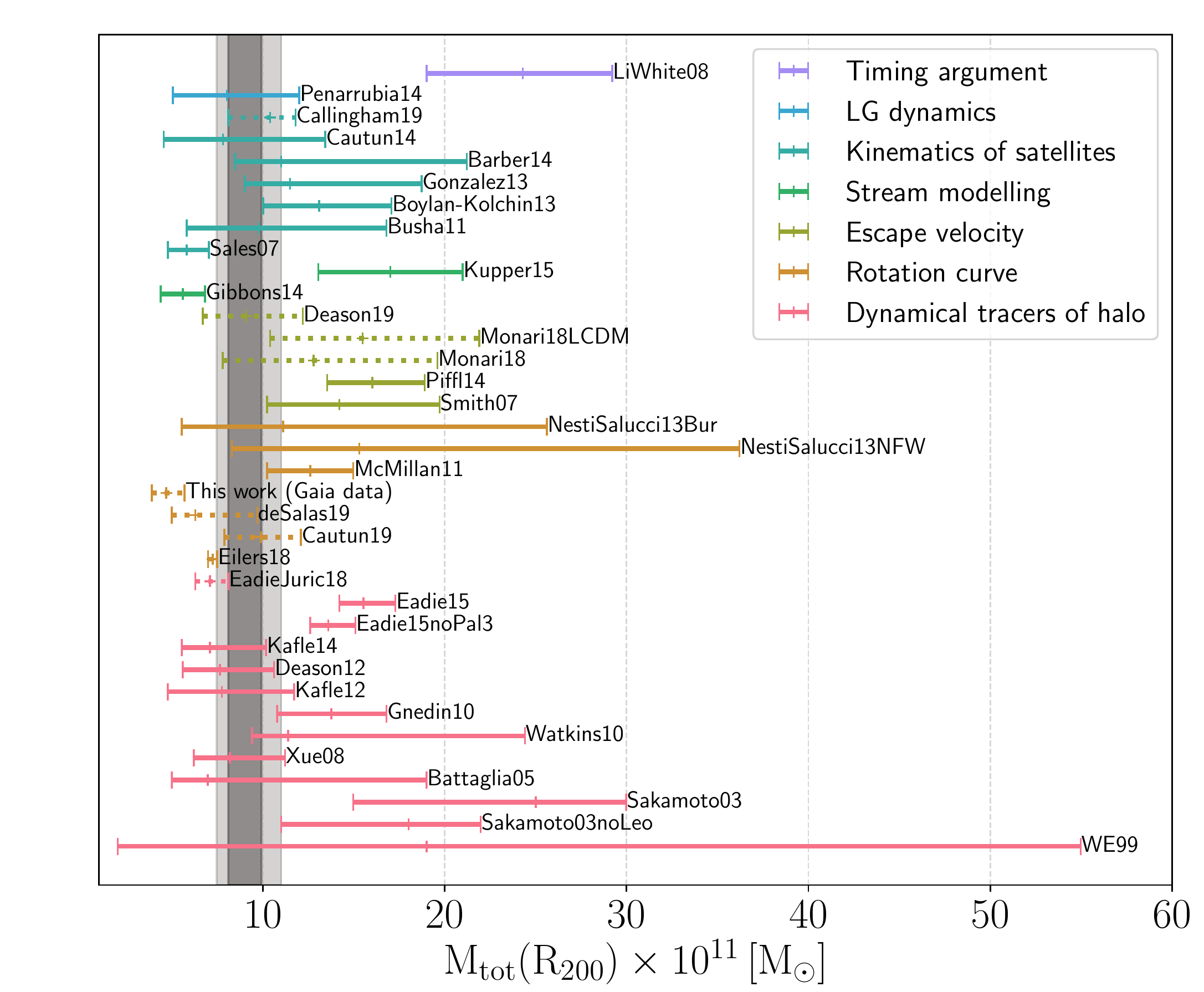}
\caption{\label{fig:mw_virial_mass_comparison} 
Comparison of our inferred Milky Way mass $M_{\rm tot}$ (dark + baryonic mass) with a selection of previous estimates based on different data sets and methodologies. 
The vertical gray shaded regions show our 68\% and 95\% credible intervals (HPD) for our Bayesian model-averaged value of $M_{\rm tot}$, see Eq.~\eqref{eqn:MtotMA}. The brown dashed errorbar referred as ``This work (Gaia data)" shows the 68\% credible interval (HPD) for our Bayesian model-averaged determination of the MW virial mass using the Gaia DR2 data from \cite{2019ApJ...871..120E}.
Solid error bars represent the halo mass estimates coming from~\cite{Li:2007eg,Penarrubia:2014oda,Cautun:2014dda,Barber:2013oua,Gonzalez:2013pqa,2013ApJ...768..140B,2011ApJ...743...40B,Sales:2007hp,Sales:2007hr,K_pper_2015,10.1093/mnras/stu1986,2014A&A...562A..91P,Smith:2006ym,Nesti:2013uwa,McMillan:2011wd,2016ApJ...829..108E,2012ApJ...761...98K,Kafle:2014xfa,Deason:2012wm,Gnedin:2010fv,Watkins:2010fe,Xue:2008se,Battaglia:2005rj,Sakamoto:2002zr}, while dashed error bars correspond to the latest measurements using Gaia data~\cite{Callingham+2019, Deason+2019, Monari+2018, Eadie&Juric2018,2019ApJ...871..120E, cautun2019milky,2019JCAP...10..037D}. 
The color coding indicates the technique used to estimate the virial mass of the Galaxy.
Quoted uncertainties correspond to 68\% confidence/credible intervals. Note that various studies may adopt different values of $R_0$ and/or $V_0$, which can introduce an apparent incompatibility.}
\end{figure}

\section{Comparison with other mass estimates}
\label{sec:compmass}

\par In figure~\ref{fig:mw_virial_mass_comparison} we compare our ``fiducial'' determination, namely our model-averaged determination of $M_{\rm tot}$, \footnote{Note that some literature adopts the definition of mass $M_{\rm tot} = 200 \rho_{cr} 4\pi/3 R_{\rm vir}^3$, slightly different from the one used throughout this work.} (Eq. \ref{eqn:M200MA}), for the \texttt{galking}$_{12}$+Huang data, and $(R_0, V_0)=(8.34\,{\rm kpc}, 239.89\,{\rm km/s})$, 
with results from previous studies.
Rather than providing a complete review of values from the literature (for which we address the reader to the recent \cite{Wang:2019ubx}), we present a representative set of estimates obtained with different techniques (shown in different colors in figure~\ref{fig:mw_virial_mass_comparison}) in order to highlight the spread in measurements of the Milky Way halo mass. 
These methods include the timing argument~\cite{Li:2007eg}, dynamics of the Local Group (LG)~\cite{Penarrubia:2014oda}, kinematics of satellites~\cite{Callingham+2019, Cautun:2014dda,Barber:2013oua,Gonzalez:2013pqa,2013ApJ...768..140B,2011ApJ...743...40B,Sales:2007hp,Sales:2007hr,2019arXiv191202086L}, modelling of stellar streams~\cite{K_pper_2015,10.1093/mnras/stu1986} and the escape velocity~\cite{Deason+2019, Monari+2018, 2014A&A...562A..91P, Smith:2006ym}, the rotation curve technique~\cite{Nesti:2013uwa,McMillan:2011wd,cautun2019milky}, and the use of kinematical tracers of the stellar halo~\cite{Eadie&Juric2018, 2016ApJ...829..108E,2012ApJ...761...98K,Kafle:2014xfa,Deason:2012wm,Gnedin:2010fv,Watkins:2010fe,Xue:2008se,Battaglia:2005rj,Sakamoto:2002zr}.
It is important to note that even those estimates that use the same technique do not always agree. In particular, different estimates of the halo mass of the Galaxy based on dynamical tracers range from $\sim 8\times 10^{11}\,\rm M_{\odot}$ to $\sim 20\times 10^{11}\,\rm M_{\odot}$ as shown in figure~\ref{fig:mw_virial_mass_comparison}.

Our halo mass estimate (vertical gray shaded regions in figure~\ref{fig:mw_virial_mass_comparison}) is at the lower end of most mass estimates in the literature. However, it is in agreement with recent mass determinations (e.g.~\cite{Penarrubia:2014oda,Cautun:2014dda,Kafle:2014xfa}), particularly with those based on the latest Gaia data (e.g.~\cite{Callingham+2019,Deason+2019,Eadie&Juric2018,2019JCAP...10..037D,2019ApJ...871..120E}).

\begin{figure}
\centering
\includegraphics[width=1\textwidth]{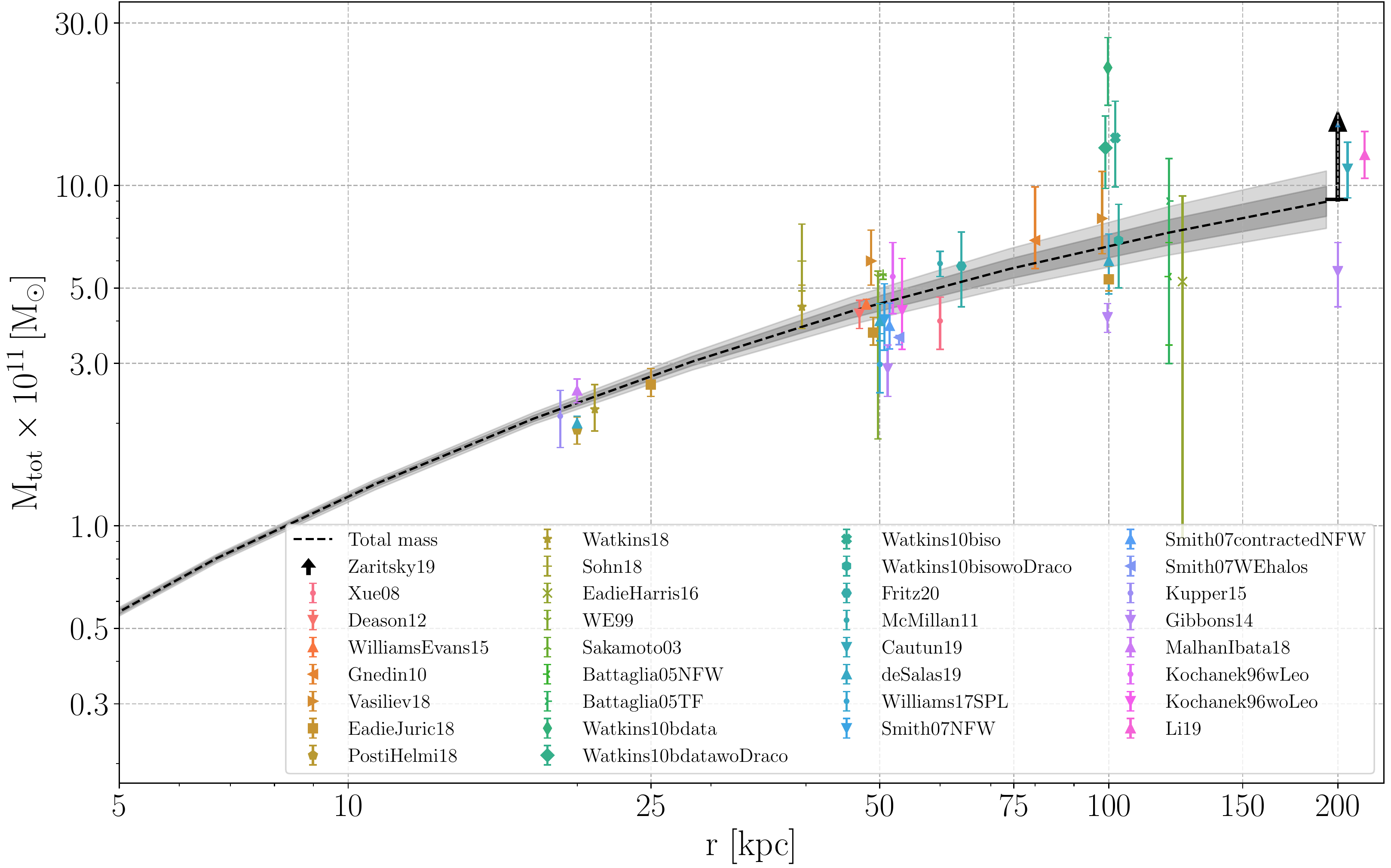}
\caption{\label{fig:mass_profiles} 
Milky Way mass profile for the maximum posterior density parameters (black dashed curve) and the corresponding 68\%/95\% credible intervals (dark/light gray shade), conditioned on the radius and model-averaged over baryonic morphologies. Also plotted are results from several other studies of the Milky Way's cumulative mass distribution~\cite{Xue:2008se,Deason:2012wm,10.1093/mnras/stv1967,2012ApJ...761...98K,Gnedin:2010fv,Dehnen:2006cm,2019MNRAS.484.2832V,Eadie&Juric2018,2019A&A...621A..56P,2019ApJ...873..118W,Sohn_2018,2016ApJ...829..108E,Sakamoto:2002zr,Battaglia:2005rj,Watkins:2010fe,McMillan:2011wd,2017MNRAS.468.2359W,K_pper_2015,10.1093/mnras/stu1986,2019MNRAS.486.2995M,Kochanek:1995xv,Wilkinson:1999hf,Smith:2006ym, cautun2019milky,2019arXiv191202086L, 2019JCAP...10..037D, 2020arXiv200102651F}. The markers around 50~kpc and 100~kpc are artificially dispersed horizontally so that they are distinguishable. 
The black arrow denotes the latest lower bound for $M_{\rm tot}$ from~\cite{Zaritsky+2019}. As with figure~\ref{fig:mw_virial_mass_comparison}, note that different choices for $R_0$ and/or $V_0$ among studies can induce apparent discrepancies.}
\end{figure}

In figure~\ref{fig:mass_profiles}, we show the total mass profile of the Milky Way as a function of galactocentric radius, and indicate 68\% and 95\% HPD regions.\footnote{The inferred enclosed mass as a function of galactocentric radius is also listed in table~\ref{tab:virial_total_mass_profile} of Appendix~\ref{App:mass profile}.} Note that the shaded region shows the uncertainty at each fixed radius.  We find that the uncertainty on the total mass increases with radius. We additionally notice an anti-correlation in the posterior between the mass of the baryons and the dark matter within a given radius due to the fact that the rotation curve is sensitive to the total mass. We have verified that, as one might expect, this anti-correlation is present in the inner 20~kpc and disappears beyond that radius, where the baryonic contribution becomes negligible. As a result, the total mass is constrained much more tightly than the individual component masses within around 20~kpc.

Figure~\ref{fig:mass_profiles} also shows estimates from previous studies of the Milky Way mass within various radii. Our total mass profile is compatible at 1$\sigma$ with most estimates summarised in the figure. The estimates that fall outside our 68\% HPD region are~\cite{Dehnen:2006cm, Sohn_2018, 10.1093/mnras/stu1986, Smith:2006ym, Eadie&Juric2018, 2019MNRAS.484.2832V, Sakamoto:2002zr, Watkins:2010fe, Battaglia:2005rj, McMillan:2011wd}, though we stress that many of these studies adopt different values for the galactic parameters, thus preventing a straightforward comparison.

\subsection{Comparison with Gaia data}
\par We compare our determination of the MW virial mass with that obtained using the Gaia DR2 catalogue. To do so we apply our procedure to the rotation curve derived from Gaia data by Eilers~{\em et~al.}~\cite{2019ApJ...871..120E}. The latter are provided as circular velocities, regressed from a Jeans analysis, for the values $(R_0, V_0)=(8.122\,{\rm kpc}, 229\,{\rm km/s})$.
We have therefore rescaled our \texttt{galkin}$_{12}$ dataset to these values (in a procedure analogous to that described in Section~\ref{subsec:V0}), and performed a model-average estimate, using both the Gaia dataset \cite{2019ApJ...871..120E}, and our \texttt{galkin}$_{12}$ data, separately, thus being able to directly compare the two determinations in a physically meaningful way. {figure~\ref{fig:rc_datasets} shows the Eilers~{\em et~al.} data, together with the {\tt galkin$_{12}$} data, rescaled to the same Galactic parameters adopted in the Eilers~{\em et~al.} analysis. We also show three best-fit models: the best-fit for the {\tt galkin}$_{12}$ data ($\chi^2$/dof = 0.6, with the best-fitting morphology being RBR); the best-fit for the Eilers~{\em et~al.} data ($\chi^2$/dof = 0.1, with the best-fitting morphology being ZCM), and the best-fit for the Eilers~{\em et~al.} data using the E2BR morphology, which approximately matches the baryonic model B2 used in~\cite{2019JCAP...10..037D} (giving $\chi^2$/dof = 1.2; the difference with respect to the ZCM morphology is mostly driven by the poorer fit to the microlensing optical depth and stellar surface density when using the E2BR morphology, rather than from a significantly different fit to the rotation curve data). 
The dark matter mass MAP estimates and their 68\% credible intervals are $M_{200}^{\rm DM}=(8.0^{+13.2}_{-3.6})\times$10$^{11}$M$_\odot$ ({\tt galkin}$_{12}$ data and RBR morphology), $M_{200}^{\rm DM}=(4.4^{+1.0}_{-0.6})\times$10$^{11}$M$_\odot$  (Eilers et al.~\cite{2019ApJ...871..120E} data and ZCM morphology), and  $M_{200}^{\rm DM}=(6.8^{+1.8}_{-1.2})\times$10$^{11}\rm M_\odot$ (Eilers et al.~\cite{2019ApJ...871..120E} data and E2BR morphology).

\begin{figure}[H]
\centering
\includegraphics[width=0.95\textwidth]{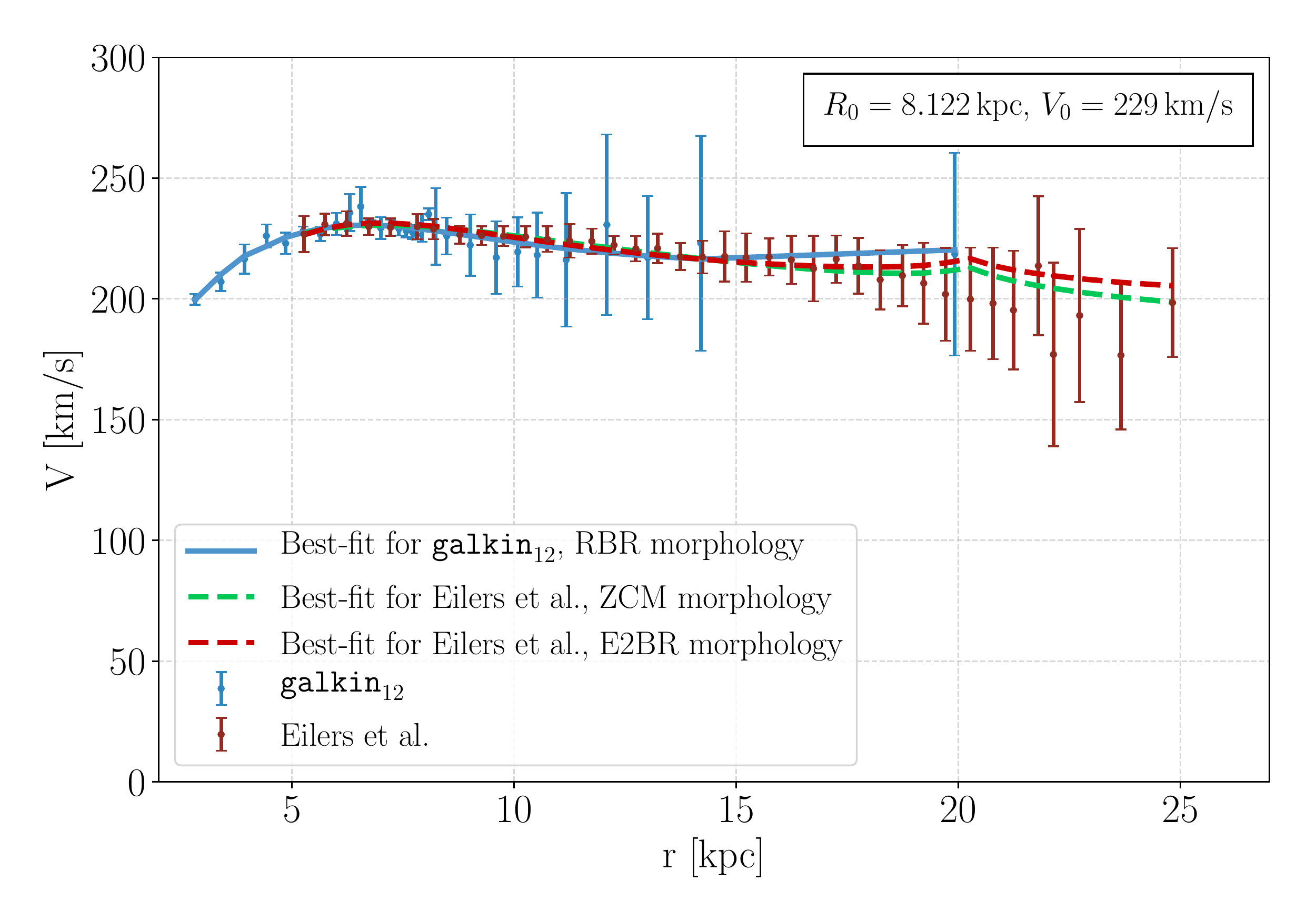}
\caption{\label{fig:rc_datasets} 
{Rotation curve data from {\tt galkin}$_{12}$ and Gaia (Eilers et al.~\cite{2019ApJ...871..120E}),  re-scaled to the Galactic parameters $R_0=8.122$ kpc and $V_0=229$ km/s, and total best-fit (baryonic + dark matter) rotation curves for three different morphologies (see text for details). 
}}
\end{figure}

For the model-averaged analysis over all morphologies, we obtain the following results: $M_{200}^{\rm DM}=(3.3^{+7.2}_{-0.8})\times$10$^{11}$M$_\odot$ (\texttt{galkin}$_{12}$ data) and 
$M_{200}^{\rm DM}=(4.7^{+1.0}_{-0.8})\times$10$^{11}$M$_\odot$ (Eilers~{\em et~al.} data), which are in agreement within the statistical uncertainties. 
{Our dark matter estimate from Gaia data is lower than the values quoted by \cite{2019ApJ...871..120E} (see figure~\ref{fig:mw_virial_mass_comparison}). This is because we use a different dark matter density profile and we allow the baryonic mass and morphology to vary. Our estimate is also somewhat lower than that of \cite{2019JCAP...10..037D}. In this case the difference can be ascribed to our model-averaged result that gives a higher weight to a different baryonic model than the ones assumed in \cite{2019JCAP...10..037D}. In particular, our E2BR morphology, which is similar to B2 model of de Salas~{\em et~al.} \cite{2019JCAP...10..037D}, is strongly downweighted in the Bayesian model averaging when compared to others. As mentioned above, when assuming the E2BR morphology we obtain a dark matter mass of  $M_{200}^{\rm DM}=(6.8^{+1.8}_{-1.2})\times$10$^{11}$M$_\odot$ which is in good agreement with the estimate of \cite{2019JCAP...10..037D} obtained for their baryonic model B2.}}

\section{Conclusions}
\label{sec:conclusions}

We have used rotation curve data to estimate the dark and total mass of the Milky Way and performed a careful assessment of the robustness of these estimates. Our Bayesian framework allows us to marginalize over nuisance parameters as well as average over baryonic morphologies, thus accounting for uncertainty in the shape of the Milky Way's distribution of baryons. We have identified a residual dependency on the assumed parameterization of the underlying dark matter density profile. Changing the adopted shape of the dark matter density profile yield a change in the inferred Milky Way virial mass $M_{200}^{\rm DM}$ by $\sim 48 \%$ (or 0.28~dex). 
The value of $M_{200}^{\rm DM}$ is also dependent on the local circular velocity $V_0$. We find that a variation of $V_0$ within the latest observational uncertainties leads to an uncertainty of 0.27~dex in $M_{200}^{\rm DM}$.
We have obtained estimates for the dark matter mass within the virial radius:  
\begin{equation*}
\log_{10}M_{200}^{\rm DM}/\mathrm{M}_{\odot}=11.92^{+0.06}_{-0.05}{\rm(stat)}\pm{0.28}\pm0.27{\rm(syst)},
\end{equation*}
and the for the total (sum of dark matter and baryons) mass within the virial radius:
\begin{equation*}
\log_{10}M_{\rm tot}/ \mathrm{M}_{\odot}=11.95^{+0.04}_{-0.04}{\rm(stat)}\pm0.25\pm0.25{\rm(syst)}.
\end{equation*}
The first systematic error comes from the choice of dark matter density profile, while the second is associated with the uncertainties on the Sun's velocity $V_0$.
As it can be seen, these mass estimates are precise from a statistical point of view, but suffer from a relatively large remaining systematic uncertainty. 

Finally, we have compared the results from our analysis of halo and total Milky Way mass with estimates based on previous studies that use different techniques and find our determination to be in agreement with most studies in the literature.
We have also applied our procedure to the Gaia DR-2 data, obtaining a determination in good agreement with that proceeding from different datasets.

\appendix
\renewcommand\thefigure{\thesection.\arabic{figure}}

\section{Mass profile constraints}
\label{App:mass profile}
In this Appendix we provide the posterior constraints on the virial and total mass profile obtained in this analysis (after model averaging). The total mass profile is plotted in figure~\ref{fig:mass_profiles} and figure~\ref{fig:mass_profiles_dif_par}.

\begin{table}[h] 
\centering
\addtolength{\tabcolsep}{6pt}
\scalebox{1}{%
\begin{tabular}{ | c | c | c |}
\hline 
$r$ [kpc] & $M_{\rm tot}$  [$10^{11}\; \rm M_{\odot}$] & $M_{\rm DM}$ [$10^{11}\; \rm M_{\odot}$] \\
\hline 
2.57 & $0.23^{+0.006\,(0.011)}_{-0.005\,(0.011)}$&  $0.10^{+0.03\,(0.05)}_{-0.02\,(0.05)}$  \\[1ex] 
4.15   & $0.45^{+0.008\,(0.014)}_{-0.007\,(0.012)}$ & $0.22^{+0.04\,(0.07)}_{-0.03\,(0.08)}$ \\[1ex] 
6.71   & $0.80^{+0.01\,(0.02)}_{-0.01\,(0.03)}$ & $0.45^{+0.05\,(0.09)}_{-0.05\,(0.11)}$ \\[1ex] 
10.85  & $1.33^{+0.02\,(0.04)}_{-0.03\,(0.05)}$ &  $0.87^{+0.06\,(0.12)}_{-0.06\,(0.12)}$ \\[1ex] 
17.53  & $2.07^{+0.04\,(0.09)}_{-0.04\,(0.08)}$ &  $1.51^{+0.07\,(0.14)}_{-0.07\,(0.13)}$ \\[1ex] 
28.33  & $3.04^{+0.10\,(0.19)}_{-0.08\,(0.17)}$ &  $2.46^{+0.08\,(0.18)}_{-0.12\,(0.21)}$ \\[1ex] 
45.79 & $4.27^{+0.22\,(0.43)}_{-0.19\,(0.37)}$ &  $3.63^{+0.23\,(0.42)}_{-0.18\,(0.36)}$ \\[1ex] 
74.0 & $5.68^{+0.40\,(0.83)}_{-0.37\,(0.65)}$ &  $5.06^{+0.38\,(0.79)}_{-0.38\,(0.67)}$ \\[1ex] 
119.57 & $7.26^{+0.66\,(1.40)}_{-0.58\,(1.03)}$ &  $6.59^{+0.69\,(1.43)}_{-0.55\,(0.97)}$ \\[1ex] 
193.24 & $8.95^{+0.98\,(2.07)}_{-0.84\,(1.48)}$ &  $8.26^{+1.21\,(2.09)}_{-0.92\,(1.43)}$ \\[1ex] 
\hline
\end{tabular}}
\caption{Constraints on the enclosed total and dark matter mass as a function of galactic radius (after model averaging). The table gives the MAP values and the 68\%\,(95\%) credible intervals, conditional on radius. The dark matter mass profile $M_{\rm DM}$ is shown in figure~\ref{fig:mass_profiles_dif_par} and the total mass profile $M_{\rm tot}$ is shown in figure~\ref{fig:mass_profiles_dif_par} and figure~\ref{fig:mass_profiles}. }
\label{tab:virial_total_mass_profile} 
\end{table}

\section{Baryonic morphologies}
\label{App:baryonic_morphologies}

A model for the baryonic component of the Milky Way is needed to constrain the Milky Way's total gravitational potential.
Despite many efforts in the amount and quality of observations, the actual distribution of visible matter in the Milky Way remains uncertain. Following the approach of \cite{Iocco:2015xga, Pato:2017yai, Iocco:2016itg}, we account for these uncertainties by taking into account different functional shapes available in the literature for both the disk(s) and bulge components.
As introduced in Section~\ref{sec:baryons}, for each combination of bulge and disk we compute the corresponding contribution to the rotation curve. 
In table~\ref{tab:summary_baryonic_morphologies} we briefly describe the bulge and disk morphologies adopted in this work. 
For further details on the derived baryonic models, we refer the interested reader to \cite{2015JCAP...12..001P,Iocco:2015xga} and the original references.

\begin{table}[H]
\centering
\addtolength{\tabcolsep}{6pt}
\renewcommand{\arraystretch}{1.1}
\scalebox{0.9}{%
\begin{tabular}{ | c | c | c | c |}
\hline
 & model & specification & Ref. \\
\rowcolor{Gray}
bulge & G2 & gaussian  &  \cite{Stanek:1995ws} \\
\rowcolor{Gray}
      & E2* & exponential & \cite{Stanek:1995ws} \\
\rowcolor{Gray}
      & V  &  truncated power law  & \cite{Vanhollebeke} \\
\rowcolor{Gray}
      & BG &  truncated power law  & \cite{Bissantz:2001wx} \\
\rowcolor{Gray}
      & Z  & gaussian plus nucleus & \cite{Zhao:1995qh} \\
\rowcolor{Gray}
      & R  &  double ellipsoid  & \cite{Robin} \\
disk  & BR   & thin plus thick & \cite{2013ApJ...779..115B} \\ 
      & HG*   & thin plus thick & \cite{2003ApJ...592..172H} \\
      & CM   & thin plus thick & \cite{2011MNRAS.416.1292C} \\
      & dJ   & thin plus thick plus halo & \cite{2010ApJ...714..663D} \\
      & J    & thin plus thick plus halo & \cite{Juric:2005zr} \\
\hline
\end{tabular}}
\caption{Summary of bulge and disk morphologies. For further details see \cite{2015JCAP...12..001P,Iocco:2015xga}.
The configuration marked with an asterisk indicates our reference morphology.}

\label{tab:summary_baryonic_morphologies} 
\end{table}

\section*{Acknowledgements}

We thank the referee for his/her comments that helped improve the paper. E.~K.'s work at ICTP-SAIFR during the first stages of this work has been supported by the S\~ao Paulo Research Foundation (FAPESP) under Grant No. 2016/26288-9. E.~K. is supported by the grant AstroCeNT: Particle Astrophysics Science and Technology Centre is carried out within the International Research Agendas programme of the Foundation for Polish Science co-financed by the European Union under the European Regional Development Fund.
M.~B. acknowledges hospitality at the Gran Sasso Science Institute (GSSI) through funding from the European Research Council (ERC) under the European Union’s Horizon 2020 research and innovation programme (grant agreement No 818744)-- for part of the duration of this work. M.~B. is supported by the ERDF Centre of Excellence project TK133.
F.~I.'s work has been partially supported by the research grant number 2017W4HA7S ``NAT-NET: Neutrino and Astroparticle Theory Network'' under the program PRIN 2017 funded by the Italian Ministero dell'Universit\`a e della Ricerca (MUR), and through a SIMONS Foundation fellowship for the first stages at ICTP--SAIFR.
AG-S, F.~I.~and R.~T.~are supported by Grant ST/N000838/1 from the Science and Technology Facilities Council (UK). R.~T. was partially supported by a Marie-Sklodowska-Curie RISE (H2020-MSCA-RISE-2015-691164) Grant provided by the European Commission.  
This work has been supported through the FAPESP/Imperial College London exchange grant ``SPRINT'', process number 2016/50006-3.
Numerical resources for this research have been supplied by the Center for Scientific Computing (NCC/GridUNESP) of the S\~ao Paulo State University (UNESP).

\clearpage
\bibliographystyle{JHEP} 
\bibliography{MWmassKarukes-V2} 

\end{document}